\documentclass[letterpaper,twocolumn,10pt]{article}
\usepackage{styles/usenix}

\newcommand{\papertitle}[0]{Terminal Brain Damage: Exposing the Graceless Degradation\\ in Deep Neural Networks Under Hardware Fault Attacks}

\usepackage{adjustbox}
\usepackage{combelow}
\usepackage{amsmath}
\usepackage{amssymb}
\usepackage{nth}
\usepackage{balance}
\usepackage{pifont}
\usepackage[inline]{enumitem}
\usepackage{threeparttable}
\usepackage[normalem]{ulem}  	
\usepackage{soul}				
\usepackage{xcolor}				
\usepackage{cite}            			
\usepackage{calc}            			
\usepackage{shadow}          		
\usepackage{fancyvrb}        		
\usepackage{xspace}          		
\usepackage{ifthen}          			
\usepackage{comment}         		
\usepackage{multirow}        		
\usepackage{dcolumn}         		
\usepackage{colortbl}
\usepackage{tabularx}        		
\usepackage{ifpdf}           			
\usepackage{rotating}        		
\usepackage{listings}
\usepackage{wrapfig}         		
\usepackage{pdfpages}			
\usepackage{graphicx}        		
\usepackage{subcaption}
\usepackage{fixltx2e}
\usepackage[hang,flushmargin,bottom]{footmisc}
\usepackage{booktabs}
\usepackage{adjustbox}
\usepackage[ruled,vlined,linesnumbered]{algorithm2e}
\usepackage{array}
\usepackage{tikz}

\usepackage{nicefrac} 
\usepackage{authblk}
\usepackage{etoolbox}

\usepackage{xcolor}
\newcommand{\hcell}[1]{
	\cellcolor[gray]{.9}#1
}

\makeatletter
\def\@copyrightspace{\relax}
\makeatother

\RequirePackage[normalem]{ulem} 
\RequirePackage{color}\definecolor{RED}{rgb}{1,0,0}\definecolor{BLUE}{rgb}{0,0,1} 

\usepackage[table, subgroups]{svn-multi} 

\svnidlong
{$HeadURL: https://mc2.umiacs.umd.edu/svn/SDSatUMD/papers/2019_dnn_fault_attacks/dnn_fault_attacks.tex $}
{$LastChangedDate: 2019-06-02 14:57:52 -0400 (Sun, 02 Jun 2019) $}
{$LastChangedRevision: 7335 $}
{$LastChangedBy: shhong $}

\def\paperversionmajor{1}          
\def\paperversionminor{\svnrev}    

\def\monthName#1{\ifcase#1\or
  January\or February\or March\or April\or May\or June\or
  July\or August\or September\or October\or November\or December\fi}

\hypersetup{pdftitle={\papertitle}}
\hypersetup{pdfproducer={v. \paperversionmajor.\paperversionminor}}
\ifpdf
\fi

\newcommand{\TODO}[2][]{\noindent\colorbox{lime}{
    \color{red}
    \parbox{\minof{.97\hsize}{\widthof{#1} * \real{1.15} + \widthof{ #2}}}{\textbf{#1} #2}
    \par}
}

\sethlcolor{lime}

\newcommand{\pit}[1]{
  \textcolor{purple}{\textbf{Pit:} #1}
}
\newcommand{\cri}[1]{
  \textcolor{blue}{\textbf{Cri:} #1}
}

\newcounter{hypothesis}                                     

\newcommand{\rh}{Rowhammer\xspace}
\newcommand{\ndimm}{12\xspace} 
\newcommand{\nRhSessions}{300\xspace}
\newcommand{\nRounds}{25\xspace}

\newcommand{\nDropDouble}{15.6\xspace}
\newcommand{\nCrash}{6\xspace}

\newcommand{\prob}{Blind\xspace}
\newcommand{\deter}{Surgical\xspace}
\newcommand{\avgWeights}{50}
\newcommand{\clkscrew}{CLK{\footnotesize{}SCREW}\xspace}
\newcommand{\xmark}{\ding{55}}
\newcommand{\cmark}{\ding{51}}

\newcommand{\PreserveBackslash}[1]{\let\temp=\\#1\let\\=\temp}
\newcolumntype{C}[1]{>{\PreserveBackslash\centering}p{#1}}
\newcolumntype{R}[1]{>{\PreserveBackslash\raggedleft}p{#1}}
\newcolumntype{L}[1]{>{\PreserveBackslash\raggedright}p{#1}}

\begin{document}

\date{}


\title{\papertitle}

\author{
	{\rm Sanghyun Hong, Pietro Frigo\textsuperscript{\textdagger}, Yi\u{g}itcan Kaya, Cristiano Giuffrida\textsuperscript{\textdagger}, Tudor Dumitra\cb{s}}\\
	\textit{University of Maryland, College Park}\\
	\textsuperscript{\textdagger}\textit{Vrije Universiteit Amsterdam}
}

\maketitle

\pagestyle{empty}

%
%

\begin{abstract}
\noindent 
%
Deep neural networks (DNNs) have been shown to tolerate ``brain damage'': cumulative changes to the network's parameters (e.g., pruning, numerical perturbations) typically result in a graceful degradation of classification accuracy. 
%
%
However, the limits of this natural resilience are not well understood in the presence of small adversarial changes to the DNN parameters' underlying memory representation, such as bit-flips that may be induced by hardware fault attacks.
%
%
We study the effects of bitwise corruptions on 19 DNN models---six architectures on three image classification tasks---and we show that most models have at least one 
parameter that, after \emph{a specific bit-flip in their bitwise representation}, 
causes an accuracy loss of over 90\%.
%
We employ simple 
heuristics to efficiently identify the parameters 
likely to be vulnerable. 
We estimate that
40--50\% of the parameters 
in a model might lead to an accuracy drop greater than 10\% when individually subjected to such single-bit perturbations.
%
To demonstrate how an adversary could take advantage of this vulnerability, we study the impact of an 
exemplary hardware fault attack, \emph{\rh}, on DNNs. Specifically, we show that a \rh-enabled attacker
co-located 
in the same physical machine can inflict significant accuracy drops (up to 99\%)
even with single bit-flip corruptions and no knowledge of the model. 
%
%
Our results expose the limits of DNNs' resilience against 
parameter perturbations induced by real-world fault attacks. We
conclude by discussing possible mitigations and future research directions towards fault attack-resilient DNNs.

\end{abstract}

%
%

\section{Introduction}
\label{sec:intro}
\newcommand{\fnOne}{\footnote{The vulnerability of a parameter requires a \emph{specific} bit in its bitwise representation to be flipped. There also might be multiple such bits in the representation that, when flipped separately, trigger the vulnerability.}}

Deep neural networks (DNNs) are known to be resilient to ``brain damage''~\cite{lecun1990optimal}: typically, cumulative changes to the network's parameters result in a graceful degradation of classification accuracy. 
This property has been harnessed in a broad range of techniques, such as network pruning~\cite{li2016pruning}, which significantly reduces the number of parameters in the network and leads to improved inference times.
%
%
%
Besides structural resilience, DNN models can tolerate slight noise in their parameters with minimal accuracy degradation~\cite{an1996effects}.
Researchers have proposed utilizing this property in defensive techniques, such as adding Gaussian noise to model parameters to strengthen DNN models against adversarial examples~\cite{zhou2018breaking}.
As a result, this natural resilience is believed to make it difficult for attackers to significantly degrade the overall accuracy by corrupting network parameters.

Recent work has explored the impact of \emph{hardware faults} on DNN models~\cite{qin2017robustness, li2017understanding, reagen2018ares}. 
Such faults can corrupt the memory storing the victim model's parameters, stress-testing DNNs' resilience to bitwise corruptions.
For example, Qin et al.~\cite{qin2017robustness}, confirming speculation from previous studies~\cite{li2016pruning,li2017understanding}, 
showed that a DNN model for CIFAR10 image classification does not lose more than 5\% accuracy when as many as 2,600 parameters out of 2.5 million are corrupted by \emph{random errors}.
%
%
However, this analysis is limited to a specific scenario and only considers
accidental errors rather than attacker-induced corruptions
by means of fault attacks. 
The widespread usage of DNNs in many mission-critical systems, such as self-driving cars or aviation~\cite{chen2015deepdriving,smolyanskiy2017toward}, requires a comprehensive understanding of the security implications of such adversarial bitwise errors.

In this paper, we explore the security properties of DNNs under bitwise errors that can be induced by practical hardware fault attacks.
%
%
%
Specifically, we ask the question: \emph{How vulnerable are DNNs to 
the atomic corruption that 
a hardware fault attacker can induce%
%
?}
%
This paper focuses on single bit-flip attacks that 
are realistic as they well-approximate the constrained memory corruption primitive of practical hardware fault attacks such as \rh~\cite{seaborn2015rh}.
%
To answer this question, we conduct a comprehensive study that characterizes the DNN model's responses to single-bit corruptions in each of its parameters.
%
%

First, we implement a systematic vulnerability analysis framework that flips each bit in a given model's parameters and measures the misclassification rates on a validation set.
Using our framework, we analyze 19 DNN models; consisting of six different architectures and their variants on three popular image classification tasks---MNIST, CIFAR10, and ImageNet.
Our experiments show that, on average, $\sim$\avgWeights\% of model parameters 
are vulnerable to single bit-flip corruptions, causing relative accuracy drops above 10\%. 
Further, all 19 DNN models contain parameters that can cause an 
accuracy drop of over 90\%\fnOne.
These show that adversarial bitwise errors can lead to a \emph{graceless degradation} of classification accuracy; exposing the limits of DNNs' resilience to numerical changes.

Our framework also allows us to characterize the vulnerability by examining the impact of 
various factors: the bit position, bit-flip direction, parameter sign, 
layer width, activation function, normalization and model 
architecture.
Our key findings include: 
\begin{enumerate*}[label=\arabic*)]
	\item the vulnerability is caused by drastic spikes in a parameter's value;  
	\item the spikes in positive parameters are more threatening, however, an activation function that allows negative outputs renders the negative parameters vulnerable as well;
	\item the number of vulnerable parameters increases proportionally as the DNN's layers get wider;
	\item two common training techniques, e.g., dropout~\cite{srivastava2014dropout} and batch normalization~\cite{ioffe2015batch}, are ineffective in preventing the massive spikes bit-flips cause; and
	\item the ratio of vulnerable parameters is almost constant across different architectures (e.g., AlexNet, VGG16, and so on).
\end{enumerate*}
Further, building on these findings, we propose 
heuristics for speeding up the analysis of vulnerable parameters in large models. 

%
%
%


Second, to understand the practical impact of this vulnerability, we use \rh~\cite{kim2014flipping} as an 
exemplary hardware fault attack. 
While a variety of hardware fault attacks are documented in literature~\cite{tang2017clkscrew, breier2018practical,liu2017fault, kim2014flipping}, \rh is particularly amenable to practical, real-world exploitation.
\rh takes advantage of a widespread vulnerability in modern DRAM modules and provides an attacker with the ability to trigger controlled memory corruptions directly from unprivileged software execution.
As a result, even a constrained \rh-enabled attacker, who only needs to perform a specific memory access pattern, can mount practical attacks in a
variety of real-world environments, including cloud~\cite{razavi2016flip,xiao2016one}, browsers~\cite{gruss2016rhjs,bosman2016dedup,seaborn2015rh,frigo2018glitch}, mobile~\cite{van2016drammer,frigo2018glitch}, and servers~\cite{lipp2018nethammer,tatar2018throwhammer}.

We analyze the feasibility of \rh attacks on DNNs by 
simulating a Machine-Learning-as-a-Service (MLaaS) scenario, where the victim and attacker VMs are co-located on the same host machine in the cloud.
The co-location causes the victim and the attacker to share the same physical memory, enabling the attacker to trigger \rh bit-flips in the victim's data~\cite{razavi2016flip,xiao2016one}. 
We focus our analysis on models with an applicable memory footprint, which can realistically be targeted by hardware fault attacks such as \rh.

Our \rh results show that in a \emph{\MakeLowercase{\deter}} attack scenario, with the capability of flipping specific bits, the attacker can reliably cause severe accuracy drops in practical settings.
Further, even in a \emph{\MakeLowercase{\prob}} attack scenario, the attacker can still mount successful attacks without any control over the memory locations of the bit-flips.
Moreover, we also reveal a potential vulnerability in the \emph{transfer learning} scenario; in which a \MakeLowercase{\deter} attack targets the parameters in the layers victim model contains in common with a public one.

Lastly, we discuss directions for viable protection mechanisms, such as reducing the number of vulnerable parameters by preventing significant changes in a parameter value. 
In particular, this can be done by 1) restricting activation magnitudes and 2) using low-precision numbers for model parameters via quantization or binarization. 
We show that, when we restrict the activations using the ReLU6 activation function, the ratio of vulnerable parameters decreases from 
47\% to 3\% in AlexNet, and also,
the accuracy drops are largely contained within 10\%. 
Moreover, quantization and binarization reduce the vulnerable parameter ratio from 50\% to 1-2\% in MNIST. 
While promising, such solutions cannot deter practical hardware fault attacks in the general case, and often require 
training the victim model from scratch; hinting that more research is required towards fault attack-resilient DNNs.

\paragraph{Contributions.} We make three contributions: 

\begin{itemize}[topsep=0em,itemsep=0em]
    \item We show DNN models are 
    more vulnerable to bit-flip corruptions than previously assumed. In particular, we show adversarial bitwise corruptions induced by hardware fault attacks can easily inflict severe \emph{indiscriminate damage} by drastically increasing or decreasing the value of a model parameter.
	\item We conduct the first comprehensive analysis of DNN models' behavior 
	against single bit-flips 
	and characterize the vulnerability that a hardware fault attack can trigger.
    \item Based on our analysis, we study the impact of practical hardware fault attacks in a representative DL scenario. 
    Our analysis shows that a \rh-enabled attacker can inflict significant accuracy drops (up to 99\%) on a victim model 
    even with constrained bit-flip corruptions and no knowledge of the model.
	%
\end{itemize}

\section{Preliminaries}
\label{sec:prelim}
%
Here, we provide an overview of the required background knowledge. 
\paragraph{Deep neural networks.} A DNN can be conceptualized as a function that takes an input and returns a prediction, i.e., the inferred \emph{label} of the input instance.
The network is composed of a sequence of layers that is individually parameterized by a set of matrices, or \emph{weights}. 
Our work focuses on feed-forward DNNs---specifically on convolutional neural networks (CNNs)---in the supervised learning setting, i.e., the weights that minimize the inference error are learned from a labeled \emph{training set}.
In a feed-forward network, each layer applies a linear transformation, defined by its weight matrix, to its input---the output of the previous layer---and a bias parameter is added optionally.
After the linear transformation, a non-linear \emph{activation function} is applied; as well as other optional layer structures, such as dropout, pooling or batch normalization. 
During training, the DNN's \emph{parameters}, i.e., the weights in each layer and in other optional structures, are updated iteratively by \emph{backpropagating} the error on the training data.
Once the network converges to an acceptable error rate or when it goes through sufficient iterations, training stops and the network, along with all its parameters, is stored as a trained network.
During testing (or inference), we load the full model into the memory and produce the prediction for a given input instance, usually not in the training data.

\paragraph{Single precision floating point numbers.} The parameters of a DNN model are usually represented as IEEE754 32-bit single-precision floating-point numbers. 
This format leverages the exponential notation and trades off the large range of possible values for reduced precision. 
For instance, the number $0.15625$ in exponential notation is represented as 
$1.25\times2^{-3}$. 
Here, $1.25$ expresses the \emph{mantissa}; whereas $-3$ is the \emph{exponent}.
The IEEE754 single-precision floating-point format defines 23 bits to store the mantissa, 8 bits for the exponent, and one bit for the sign of the value. 
The fact that different bits have different influence on the represented value makes this format interesting from an adversarial perspective.
For instance, continuing or example, flipping the \nth{16} bit in the mantissa increases the value from $0.15625$ to $0.15625828$; hence, a usually negligible perturbation. 
On the other hand, a flipping the highest exponent bit would turn the value into $1.25\times2^{125}$. 
Although both of these rely on the same bit corruption primitive, they yield vastly different results.
In Sec~\ref{sec:dnn-vulnerability}, we analyze how this might lead to a vulnerability when a DNN's parameters are corrupted via single bit-flips.

\paragraph{\rh attacks.} \rh is the most common instance of software-induced fault attacks~\cite{seaborn2015rh,van2016drammer,razavi2016flip,xiao2016one,gruss2016rhjs,bosman2016dedup,frigo2018glitch, tatar2018throwhammer}. 
This vulnerability provides an aggressor with a single-bit corruption primitive at DRAM level; thus, it is an ideal attack for the purpose of our analysis.
\rh is a remarkably versatile fault attack since it only requires an attacker to be able to access content in DRAM; an ubiquitous feature of every modern system. 
By simply carrying out specific memory access patterns---which we explain in Sec~\ref{sec:exploit-rowhammer}---the attacker is able to cause extreme stress on other memory locations triggering faults on other stored data.
\section{Threat Model}
\label{sec:threat-model}


Prior research has extensively validated a DNN's resilience to parameter changes~\cite{lecun1990optimal, li2016pruning, an1996effects, zhou2018breaking, qin2017robustness, li2017understanding}, by considering random or deliberate perturbations. 
However, from a security perspective, these results provide only limited insights as they study a network's expected performance under cumulative changes.
In contrast, towards a successful and feasible attack, an adversary is usually interested in inflicting the worst-case damage under minimal changes.

We consider a class of modifications that an adversary, using hardware fault attacks, can induce in practice. 
%
We assume a cloud environment where the victim's deep learning system is deployed inside a VM---or a container---to serve the requests of external users. 
For making test-time inferences, the trained DNN model and its parameters are loaded into the system's (shared) memory and remain constant in normal operation. 
Recent studies describe this as a typical scenario in MLaaS~\cite{tramer2016stealing}. 

To understand the DNNs' vulnerability in this setting, we consider the atomic change that an adversary may induce---the single bit-flip---and we, in Sec~\ref{sec:dnn-vulnerability}, systematically characterize the damage such change may cause.
We then, in Sec~\ref{sec:exploit-rowhammer}, investigate the feasibility of inducing this damage in practice, by considering adversaries with different capabilities and levels of knowledge.

\paragraph{Capabilities.}
We consider an attacker \emph{co-located} in the same physical host machine as the victim's deep learning system. 
The attacker, due to co-location, can take advantage of a well-known software-induced fault attack, \rh~\cite{xiao2016one,razavi2016flip}, for corrupting the victim model stored in DRAM.
We take into account two possible scenarios:
\begin{enumerate*}[label=\arabic*)]
	\item a \MakeLowercase{\emph{\deter}} attack scenario where the attacker can cause a bit-flip at an intended location in the 
	victim's process memory by leveraging advanced memory massaging primitives~\cite{razavi2016flip,van2016drammer} to obtain more 
	precise results; and 
	%
	%
	\item a \emph{\MakeLowercase{\prob}} attack where the attacker lacks fine-grained control over the bit-flips; thus, is completely unaware of where a bit-flip lands in the layout of the model.
	%
\end{enumerate*}

\paragraph{Knowledge.} 
Using the existing terminology, we consider two levels for the attacker's knowledge of the victim model, e.g., the model's architecture and its parameters as well as their placement in memory:
%
\begin{enumerate*}[label=\arabic*)]
	\item a \emph{black-box} setting where the attacker has no knowledge of the victim model. Here, both the \MakeLowercase{\deter} and \MakeLowercase{\prob} attackers only hope to trigger an accuracy drop as they cannot anticipate what the impact of their bit-flips would be; and
	\item a \emph{white-box} setting where the attacker knows the victim model, at least partially. 
	Here, the \MakeLowercase{\deter} attacker can deliberately tune the attack's inflicted accuracy drop---from minor to catastrophic damage.
	Optionally, the attacker can force the victim model to misclassify a specific input sample without significantly damaging the overall accuracy.
	%
	However, the \MakeLowercase{\prob} attacker gains no significant advantage over the black-box scenario as the lack of capability prevents the attacker from acting on the knowledge. 
\end{enumerate*}
%
%

%
%

\section{Single-Bit Corruptions on DNNs}
\label{sec:dnn-vulnerability}


In this section, we 
expose DNNs' vulnerability to 
single bit-flips.
We start with an overview of our experimental setup and methodology.
We then present our findings on DNNs' vulnerability to single bit corruptions.
For characterizing the vulnerability, we analyze the impact of 
\begin{enumerate*}[label=\arabic*)]
	\item the bitwise representation of the corrupted parameter, and
	\item various DNN properties
\end{enumerate*};
on the resulting \emph{indiscriminate damage}\footnote{We use this term to indicate the severe overall accuracy drop in the model.}.
We also discuss the broader implications of the vulnerability for both the \MakeLowercase{\prob} and \MakeLowercase{\deter} attackers.
Finally, we turn our attention to two distinct attack scenarios single bit-flips lead to.

\subsection{Experimental Setup and Methodology}
\label{subsec:experimental-setup}
Our vulnerability analysis framework systematically flips the bits in a model, individually, and quantifies the impact using the metrics we define.
We implement the framework using Python 3.7\footnote{\url{https://www.python.org}} and PyTorch 1.0\footnote{\url{https://pytorch.org}} that supports CUDA 9.0 for accelerating computations by using GPUs. 
Our experiments run on the high performance computing cluster that has 488 nodes, where each is equipped with Intel 
E5-2680v2 2.8GHz 20-core processors, 180 GB of RAM, and 40 of which have 2 Nvidia Tesla K20m GPUs.
We achieve a significant amount of speed-up by leveraging a parameter-level parallelism.

\paragraph{Datasets.} 
We use three popular image classification datasets: MNIST~\cite{lecun1998gradient}, CIFAR10~\cite{krizhevsky2009learning}, and ImageNet~\cite{russakovsky2015imagenet}. 
MNIST is a grayscale image dataset used for handwritten digits (zero to nine) recognition, containing 60,000 training and 10,000 validation images of 28x28 pixels. 
CIFAR10 and ImageNet are colored image datasets used for object recognition. 
CIFAR10 includes 32x32 pixels, colored natural images of 10 classes, containing 50,000 training and 10,000 validation images. 
For ImageNet, we use the ILSVRC-2012 subset~\cite{ilsvrc2012}, resized at 224x224 pixels, composed of 1,281,167 training and 50,000 validation images from 1,000 classes.

\paragraph{Models.} 
We conduct our analysis on 19 different DNN models. 
For MNIST, we define a baseline architecture, Base (B), and generate four variants with different layer configurations: B-Wide, B-PReLU, B-Dropout, and B-DP-Norm. 
We also examine well-known LeNet5 (L5)~\cite{lecun1998gradient} and test two variants of it: L5-Dropout and L5-D-Norm. 
For CIFAR10, we employ the architecture from~\cite{suciu2018fail} as a baseline and experiment on its three variants: B-Slim, B-Dropout and B-D-Norm. 
In the following sections, we discuss why we generate these variants.
In Appendix~\ref{appendix:network-architectures}, we describe the details of these custom architectures; in Appendix~\ref{appendix:dataset-stats-hyperparams}, we present the hyper-parameters. 
For CIFAR10, we also employ two off-the-shelf network architectures: AlexNet~\cite{krizhevsky2012imagenet} and VGG16~\cite{simonyan2014very}. 
For ImageNet, we use five well-known DNNs to understand the vulnerability of large models: AlexNet, VGG16, ResNet50~\cite{he2016deep}, DenseNet161~\cite{iandola2014densenet} and InceptionV3~\cite{szegedy2016rethinking}\footnote{The pre-trained ImageNet models we use are available at: \url{https://pytorch.org/docs/stable/torchvision/models.html}.}.

\paragraph{Metrics.} To quantify the indiscriminate damage of single bit-flips, we define the Relative Accuracy Drop as $\text{RAD} = \nicefrac{(\text{Acc}_{pristine} - \text{Acc}_{corrupted})}{\text{Acc}_{pristine}}$; where $\text{Acc}_{pristine}$ and $\text{Acc}_{corrupted}$ denote the 
classification accuracies of the pristine and the corrupted models, respectively.
In our experiments, we use [RAD$>0.1$] as the criterion for 
indiscriminate damage on the model.
We also measure the accuracy of each class in the validation set to analyze whether a single bit-flip causes a subset of classes to dominate the rest. 
In MNIST and CIFAR10, we simply compute the Top-1 accuracy on the test data (as a percentage) and use the accuracy for analysis. 
For ImageNet, we consider both the Top-1 and Top-5 accuracy; however, for the sake of comparability, we report only Top-1 accuracy in Table~\ref{tbl:evaluate-flips}.
We consider a parameter as \emph{vulnerable} if it, in its bitwise representation, contains \emph{at least one bit} that triggers severe indiscriminate damage when flipped. 
For quantifying the vulnerability of a model, we simply count the number of these vulnerable parameters.


\begin{table*}[t]
\centering
	\begin{threeparttable}
	\adjustbox{width=\linewidth}{
	\begin{tabular}{@{}cL{2cm}ccC{2.6cm}C{2.6cm}C{2.6cm}C{1.6cm}C{1.6cm}@{}}
		\toprule
		\multirow{2}{*}{\textbf{Dataset}} & \multicolumn{1}{c}{\multirow{2}{*}{\textbf{Network}}} & \multirow{2}{*}{\textbf{Base acc.}} & \multirow{2}{*}{\textbf{\# Params}} & \multicolumn{3}{c}{\textbf{Speed-up heuristics}} & \multicolumn{2}{c}{\textbf{Vulnerablility}} \\ \cmidrule(l){5-7}\cmidrule(l){8-9} 
		& \multicolumn{1}{c}{} &  &  & \textbf{SV} & \textbf{SB} & \textbf{SP} & \textbf{\# Params} & \textbf{Ratio} \\ \midrule \midrule
		\parbox[t]{2mm}{\multirow{8}{*}{\rotatebox[origin=c]{90}{\textbf{MNIST}}}}
		& B(ase) & 95.71 & 21,840 & \xmark & \xmark & \xmark & 10,972 & 50.24\% \\
		& B-Wide & 98.46 & 85,670 & \xmark & \xmark & \xmark & 42,812 & 49.97\% \\
		& B-PReLU & 98.13 & 21,843 & \xmark & \xmark & \xmark & 21,663 & 99.18\% \\
		& B-Dropout & 96.86 & 21,840 & \xmark & \xmark & \xmark & 10,770 & 49.35\% \\
		& B-DP-Norm & 97.97 & 21,962 & \xmark & \xmark & \xmark & 11,195 & 50.97\% \\ \cmidrule(l){2-9}
		& L5 & 98.81 & 61,706 & \xmark & \xmark & \xmark & 28,879 & 46.80\% \\
		& L5-Dropout & 98.72 & 61,706 & \xmark & \xmark & \xmark & 27,806 & 45.06\% \\
		& L5-D-Norm & 99.05 & 62,598 & \xmark & \xmark & \xmark & 30,686 & 49.02\% \\ \midrule \midrule
		\parbox[t]{2mm}{\multirow{6}{*}{\rotatebox[origin=c]{90}{\textbf{CIFAR10}}}}
		& B(ase) & 83.74 & 776,394 		& \cmark (83.74) & \cmark (exp.) & \xmark & 363,630 & 46.84\% \\
		& B-Slim & 82.19 & 197,726 		& \cmark (82.60) & \cmark (exp.) & \xmark & 92,058 & 46.68\% \\
		& B-Dropout & 81.18 & 776,394 	& \cmark (80.70) & \cmark (exp.) & \xmark & 314,745 & 40.54\% \\
		& B-D-Norm & 80.17 & 777,806 	& \cmark (80.17) & \cmark (exp.) & \xmark & 357,448 & 45.96\% \\ \cmidrule(l){2-9}
		& AlexNet & 83.96 & 2,506,570 	& \cmark (85.00) & \cmark (exp.) & \xmark & 1,185,957 & 47.31\% \\
		& VGG16 & 91.34 & 14,736,727 	& \cmark (91.34) & \cmark (exp.) & \xmark & 6,812,359 & 46.23\% \\ \midrule \midrule
		\parbox[t]{2mm}{\multirow{5}{*}{\rotatebox[origin=c]{90}{\textbf{ImageNet}}}}
		& AlexNet & 56.52 / 79.07 & 61,100,840 & \cmark (51.12 / 75.66) & \cmark (\nth{31} bit) & \cmark (20,000) 		& 9,467\textsuperscript{\scriptsize\,SP} & 47.34\% \\
		& VGG16 & 79.52 / 90.38 & 138,357,544 & \cmark (64.28 / 86.56) & \cmark (\nth{31} bit) & \cmark (20,000) 		& 8,414\textsuperscript{\scriptsize\,SP} & 42.07\% \\
		& ResNet50 & 76.13 / 92.86 & 25,610,152 & \cmark (69.76 / 89.86) & \cmark (\nth{31} bit) & \cmark (20,000) 		& 9,565\textsuperscript{\scriptsize\,SP} & 47.82\% \\
		& DenseNet161 & 77.13 / 93.56 & 28,900,936 & \cmark (72.48 / 90.94) & \cmark (\nth{31} bit) & \cmark (20,000) 	& 9,790\textsuperscript{\scriptsize\,SP} & 48.95\% \\
		& InceptionV3 & 69.54 / 88.65 & 27,197,488 & \cmark (65.74 / 86.24) & \cmark (\nth{31} bit) & \cmark (20,000) 	& 8,161\textsuperscript{\scriptsize\,SP} & 40.84\% \\ \bottomrule
	\end{tabular}
	}
{
	\begin{tablenotes}[para]
		\item \small SV = Sampled Validation set
		\item \small SB = Specific Bits
		\item \small SP = Sampled Parameters set
	\end{tablenotes}
}
\end{threeparttable}
\caption{\textbf{Indiscriminate damages to 19 DNN models caused by single bit-flips.}}
\label{tbl:evaluate-flips}
\end{table*}


\paragraph{Methodology.} 
On our 8 MNIST models, we carry out a complete analysis: we flip each bit in all parameters of a model, in both directions---(0$\rightarrow$1) and (1$\rightarrow$0)---and compute the RAD over the entire validation set. 
However, a complete analysis of the larger models requires infeasible computational time---the
VGG16 model for ImageNet with 138M parameters would take $\approx$ 942 days on our setup. 
Therefore, based on our initial results, we devise three speed-up heuristics that aid the analysis of CIFAR10 and ImageNet models.

\paragraph{Speed-up heuristics.} 
The following three heuristics allow us to feasibly and accurately estimate the vulnerability in larger models:  \\

\vspace{-0.05in}

\noindent $\bullet$ \emph{Sampled validation set (SV).} 
After a bit-flip, deciding whether the bit-flip leads to a vulnerability [RAD$>0.1$] requires testing the corrupted model on the validation set; which might be cost prohibitive. 
This heuristic says that we can still estimate the model accuracy---and the RAD---on a sizable subset of the validation set. 
Thus, we randomly sample 10\% the instances from each class in the respective validation sets, in both CIFAR10 and ImageNet experiments.\\

\vspace{-0.1in}

\noindent $\bullet$ \emph{Inspect only specific bits (SB).} 
In Sec~\ref{sec:prelim}, we showed how flipping different bits of a IEEE754 floating-point number results in vastly different outcomes.
Our the initial MNIST analysis in Sec~\ref{subsec:bw_characterization} shows that mainly the exponent bits lead to perturbations strong enough to cause indiscriminate damage. 
This observation is the basis of our \emph{SB} heuristic that tells us to examine the effects of flipping only the exponent bits for CIFAR10 models.
For ImageNet models, we use a stronger SB heuristic and only inspect the most significant exponent bit of a parameter to achieve a greater speed-up.
This heuristic causes us to miss the vulnerability the remaining bits might lead to, therefore, its results can be interpreted as a conservative estimate of the actual number of vulnerable parameters. \\

\vspace{-0.1in}

\noindent $\bullet$ \emph{Sampled parameters (SP) set.} 
Our MNIST analysis also reveals that almost 50\% of all parameters are vulnerable to bit-flips.
This leads to our third heuristic: uniformly sampling from the parameters of a model would still yield an accurate estimation of the vulnerability.
We utilize the SP heuristic for ImageNet models and uniformly sample a fixed number of parameters---20,000---from all parameters in a model.
In our experiments, we perform this sampling five times and report the average vulnerability across all runs.
Uniform sampling also reflects the fact that a black-box attacker has a uniform probability of 
corrupting any parameter.

\subsection{Quantifying the Vulnerability That Leads to Indiscriminate Damage}
\label{subsec:indiscriminate-damage}

Table~\ref{tbl:evaluate-flips} presents the results of our experiments on single-bit corruptions, for 19 different DNN models. 
We reveal that an attacker, armed with a single bit-flip attack primitive, can successfully cause indiscriminate damage [RAD$>0.1$] and that the ratio of vulnerable parameters in a model varies between 40\% to 99\%; depending on the model.
The consistency between MNIST experiments, in which we examine every possible bit-flip, and the rest, in which we heuristically examine only a subset, shows that, in a DNN model, approximately half of the parameters are vulnerable to single bit-flips.
Our experiments also show small variability in the chances of a successful attack---indicated by the ratio of vulnerable parameters. 
With 40\% vulnerable parameters, the InceptionV3 model is the most apparent outlier among the other ImageNet models; compared to 42-49\% for the rest.
We define the vulnerability based on [RAD$>0.1$] and, in Appendix~\ref{append:rads-with-single-bitflips}, we also present how vulnerability changes within the range [$0.1 \leq$ RAD$\leq 1$].
In the following subsections, we characterize the vulnerability in relation to various factors and discuss our results in more detail.

\subsection{Characterizing the Vulnerability: Bitwise Representation}
\label{subsec:bw_characterization}
Here, we characterize the interaction how the features of a parameter's bitwise representation govern its vulnerability.


\begin{figure}[ht]
	\centering
	\includegraphics[width=\linewidth]{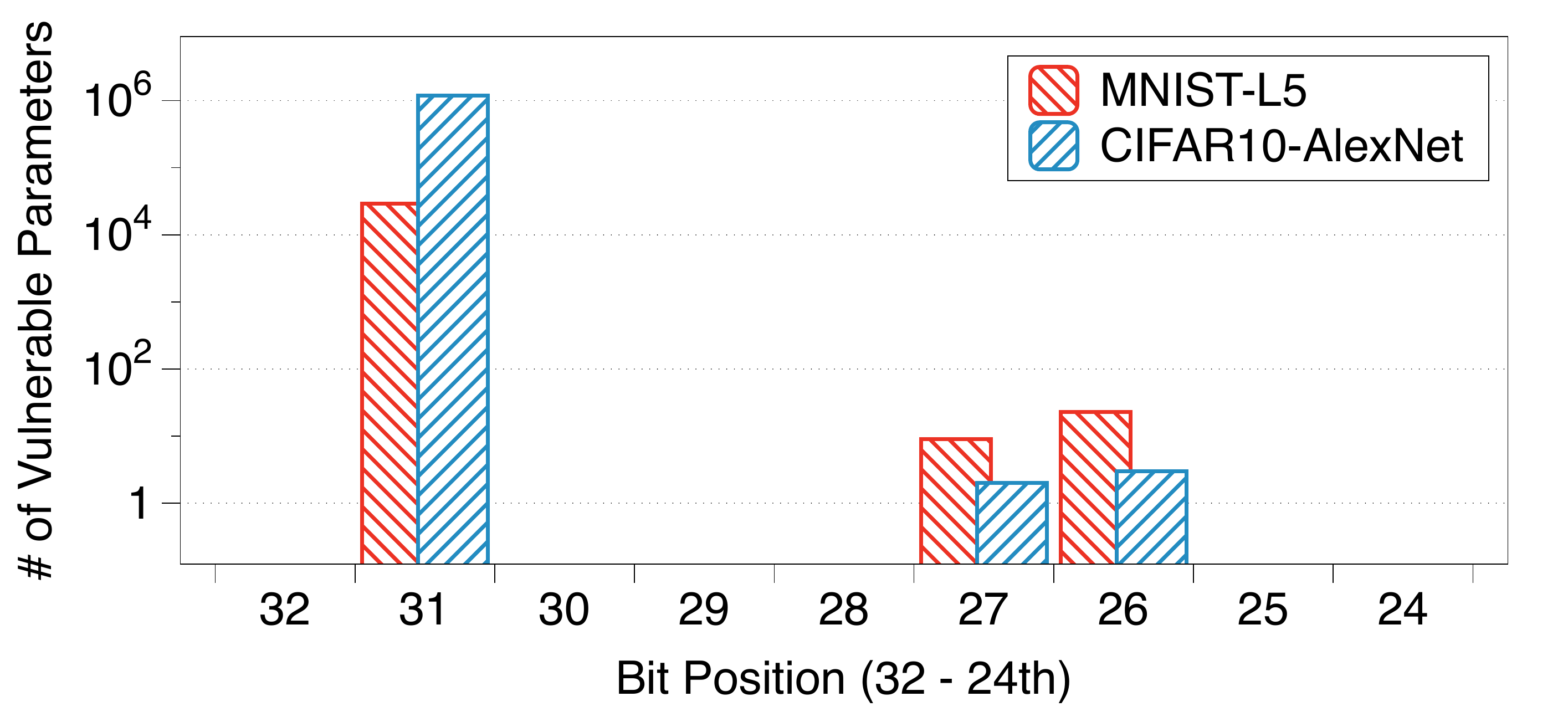}
	\caption{\textbf{The impact of the bit position.} The number of vulnerable parameters in bit positions \nth{32} to \nth{24}.}
	\label{fig:char-bitposition}
	\vspace{-1.2em}
\end{figure}

\paragraph{Impact of the bit-flip position.} 
To examine how much change in a parameter's value leads to indiscriminate damage, we focus on the position of the corrupted bits.
In Figure~\ref{fig:char-bitposition}, for each bit position, we present the number of bits---in the log-scale---that cause indiscriminate damage when flipped, on MNIST-L5 and CIFAR10-AlexNet models.
In our MNIST experiments, we examine all bit positions and we observe that bit positions other than the exponents mostly do not lead to significant damage; therefore, we only consider the exponent bits.
%
%
We find that \emph{the exponent bits, especially the \nth{31}-bit, lead to indiscriminate damage}. 
The reason is that a bit-flip in the exponents causes to a drastic change of a parameter value, whereas a flip in the mantissa only increases or decreases the value by a small amount---$[0,1]$.
We also observe that flipping the \nth{30} to \nth{28} bits is mostly inconsequential as these bits, in the IEEE754 representation, are already set to one for most values a DNN parameter usually takes---[$3.0517\times10^{-5}$, 2].


\begin{table}[b]
\vspace{-0.6em}		
\centering
\adjustbox{width=\linewidth}{
	\begin{tabular}{@{}cccccc@{}}
		\toprule
		\textbf{Direction} & \multicolumn{5}{c}{\textbf{Models (M: MNIST, C: CIFAR10)}} \\  \cmidrule(l){2-6}
		\textit{(32-\nth{24} bits)} & \textbf{M-B} & \textbf{M-PReLU} & \textbf{M-L5} & \textbf{C-B} & \textbf{C-AlexNet} \\
		\midrule
		\midrule
		(0$\rightarrow$1) & 11,019 & 21,711 & 28,902 & 314,768 & 1,185,964 \\
		(1$\rightarrow$0) & 0 & 0 & 0 & 0 & 0 \\
		\midrule
		Total & 11,019 & 21,711 & 28,902 & 314,768 & 1,185,964 \\ \bottomrule
	\end{tabular}
}
\caption{\textbf{The impact of the flip direction.} The number of effective bit-flips in 3 MNIST and 2 CIFAR10 models.}
\label{tbl:char-flipdirection}
\end{table}
\paragraph{Impact of the flip direction.} 
We answer which direction of the bit-flip, (0$\rightarrow$1) or (1$\rightarrow$0), leads to greater indiscriminate damage. 
In Table~\ref{tbl:char-flipdirection}, we report the number of effective bit-flips, i.e., those that inflict [RAD > $0.1$] for each direction, on 3 MNIST and 2 CIFAR10 models. 
We observe that \emph{only (0$\rightarrow$1) flips cause indiscriminate damage} and \emph{no (1$\rightarrow$0) flip leads to vulnerability}. 
The reason is that a (1$\rightarrow$0) flip can only decrease a parameter's value, unlike a (0$\rightarrow$1) flip.
The values of model parameters are usually normally distributed---$N(0,1)$---that places most of the values within [-1,1] range. 
Therefore, a (1$\rightarrow$0) flip, in the 
exponents, can decrease the magnitude of a typical parameter at most by one; which is not a strong enough change to inflict critical damage.
Similarly, in the sign bit, both (0$\rightarrow$1) and (1$\rightarrow$0) flips cannot cause severe damage because they change the magnitude of a parameter at most by two.
On the other hand, a (0$\rightarrow$1) flip, in the exponents, can increase the parameter value significantly; thus, during the forward-pass, the extreme neuron activation 
caused by the corrupted parameter overrides the rest of the activations.

\begin{figure*}[ht]
	\centering
	\includegraphics[width=\linewidth]{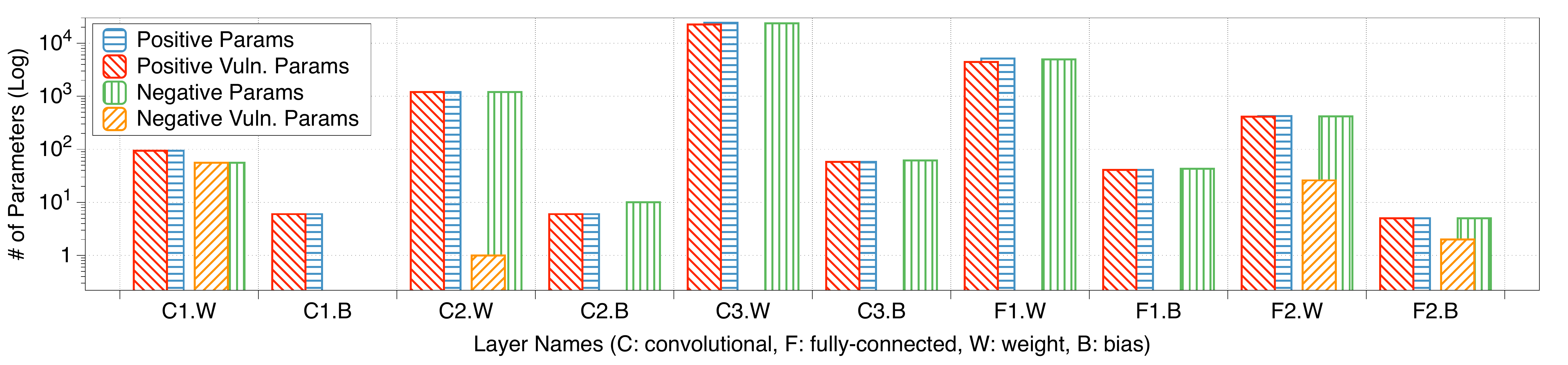}
	\caption{\textbf{The impact of the parameter sign.} The number of vulnerable positive and negative parameters, in each layer of MNIST-L5.}
	\label{fig:char-paramsign}
	\vspace{-1.0em}
\end{figure*}

\begin{figure*}[h]
	\centering
	\includegraphics[width=\linewidth]{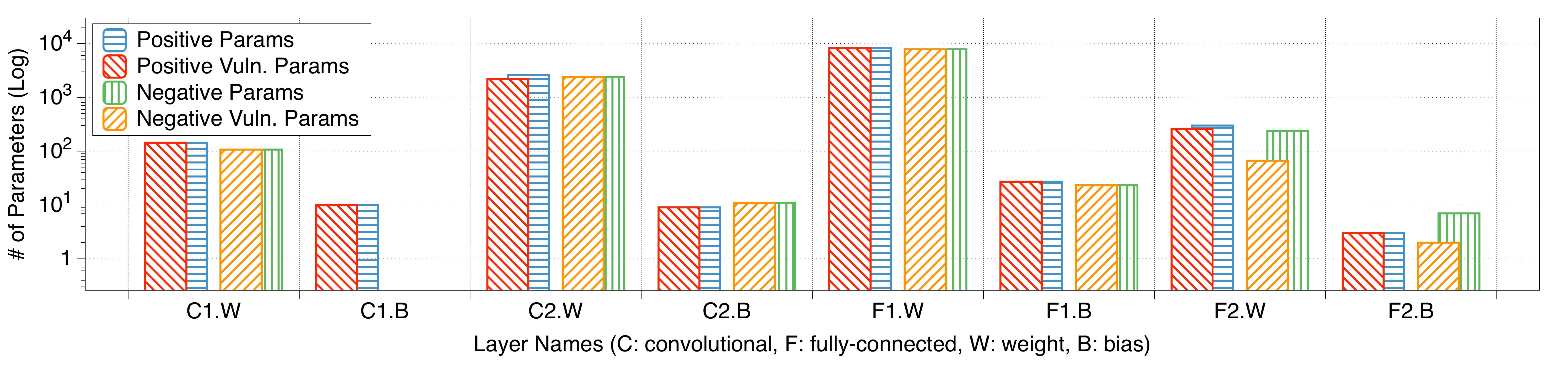}
	\caption{\textbf{The impact of the activation function.} The number of vulnerable positive and negative parameters, in each layer of MNIST-PReLU.}
	\label{fig:char-activation}
	\vspace{-1.0em}
\end{figure*}

\paragraph{Impact of the parameter sign.} 
As our third feature, we investigate whether the sign---positive or negative---of the corrupted parameter impacts the vulnerability.
In Figure~\ref{fig:char-paramsign}, we examine the MNIST-L5 model and present the number of vulnerable positive and negative parameters in each layer---in the log-scale.
Our results suggest that \emph{positive parameters are more vulnerable to single bit-flips than negative parameters}.
We identify the common ReLU activation function as the reason: ReLU immediately zeroes out the negative activation values, which are usually caused by the negative parameters.
As a result, the detrimental effects of corrupting a negative parameter fail to propagate further in the model.
Moreover, we observe that \emph{in the first and last layers, the negative parameters, as well as the positive ones, are vulnerable}.
We hypothesize that, in the first convolutional layer, changes in the parameters yield a similar effect to corrupting the model inputs directly. 
On the other hand, in their last layers, DNNs usually have the Softmax function that does not have the same zeroing-out effect as ReLU.

\subsection{Characterizing the Vulnerability: DNN Properties}
\label{subsec:char-model}
We continue our analysis by investigating how various properties of a DNN model affect the model's vulnerability to single bit-flips.

\paragraph{Impact of the layer width.} 
We start our analysis by asking whether increasing the width of a DNN affects the number of vulnerable parameters. 
In Table~\ref{tbl:evaluate-flips}, in terms of the number of vulnerable parameters, we compare the MNIST-B model with the MNIST-B-Wide model.
In the wide model, all the convolutional and fully-connected layers are twice as wide as the corresponding layer in the base model.
We see that the ratio of vulnerable parameters is almost the same for both models: 50.2\% vs 50.0\%.
Further, experiments on the CIFAR10-B-Slim and CIFAR10-B---twice as wide as the slim model---produce consistent results: 46.7\% and 46.8\%.
We conclude that \emph{the number of vulnerable parameters grows proportionally with the DNN's width} and, as a result, \emph{the ratio of vulnerable parameters remains constant at around 50\%}.

\paragraph{Impact of the activation function.} 
Next, we explore whether the choice of activation function affects the vulnerability. 
Previously, we showed that ReLU can neutralize the effects of large negative parameters caused by a bit-flip; thus, we experiment on different activation functions that allow negative outputs, e.g., PReLU~\cite{he2015delving}, LeakyReLU, or RReLU~\cite{xu2015empirical}.
These ReLU variants have been shown to improve the training performance and the accuracy of a DNN.
In this experiment, we train the MNIST-B-PReLU model; which is exactly the same as the MNIST-B model, except that it replaces ReLU with PReLU.
Figure~\ref{fig:char-activation} presents the layer-wise number of vulnerable positive and negative parameters in MNIST-B-PReLU.
We observe that \emph{using PReLU causes the negative parameters to become vulnerable and, as a result, leads to a DNN approximately twice as vulnerable as the one that uses ReLU}---50.2\% vs. 99.2\% vulnerable parameters.

\begin{figure}[h]
	\centering
	\includegraphics[width=\linewidth]{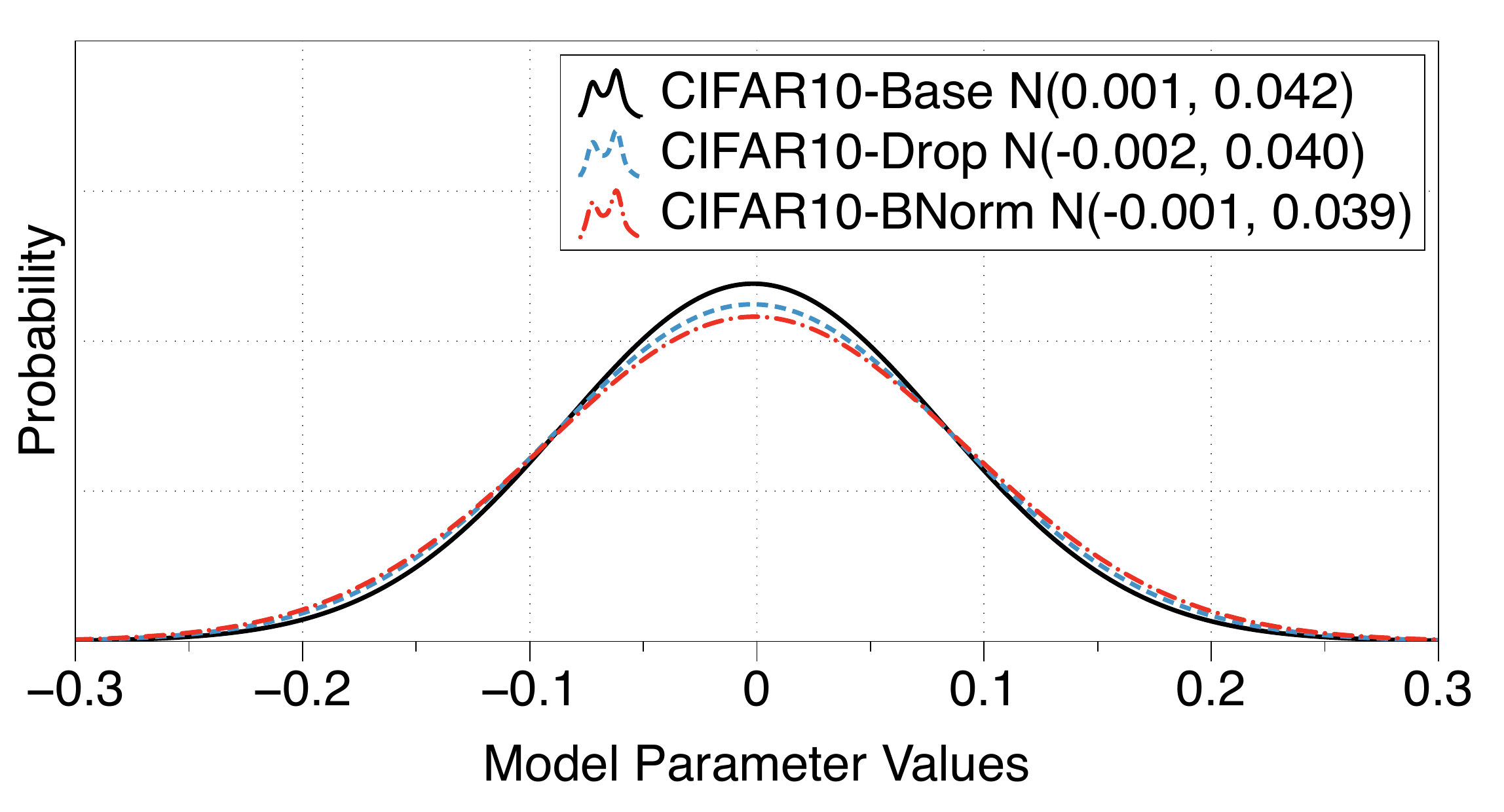}
	\caption{\textbf{The impact of the dropout and batch normalization.} The distributions of the parameter values of three CIFAR10 models variants.}
	\label{fig:char-dropbnorm}
	\vspace{-1.0em}
\end{figure}

\begin{figure*}[ht]
	\centering
	\includegraphics[width=\linewidth]{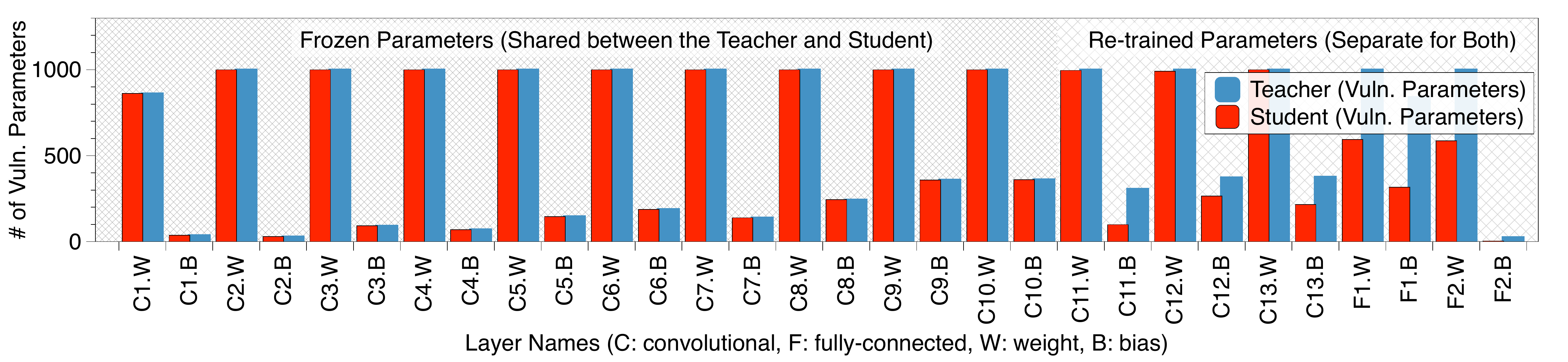}
	\caption{\textbf{The security threat in a transfer learning scenario.} The victim model---student---that is trained by transfer learning is vulnerable to the \MakeLowercase{\deter} attacker, who can see the parameters the victim has in common with the teacher model.}
	\label{fig:deterministic}
\end{figure*}

\paragraph{Impact of dropout and batch normalization.}
We confirmed that successful bit-flip attacks increase a parameter's value drastically to cause indiscriminate damage.
In consequence, we hypothesize that common techniques that tend to constrain the model parameter values to improve the performance, e.g., dropout~\cite{srivastava2014dropout} or batch normalization~\cite{ioffe2015batch}, would result in a model more resilient to single bit-flips.
Besides the base CIFAR10 and MNIST models, we train the B-Dropout and B-DNorm models for comparison.
In B-Dropout models, we apply dropout before and after the first fully-connected layers; in B-DNorm models, in addition to dropout, we also apply batch normalization after each convolutional layer.
In Figure~\ref{fig:char-dropbnorm}, we compare our three CIFAR10 models and show how dropout and batch normalization have the effect of reducing the parameter values.  
However, when we look into the vulnerability of these models, we surprisingly find that \emph{the vulnerability is mostly persistent regardless of dropout or batch normalization}---with at most 6.3\% reduction in vulnerable parameter ratio over the base network.

\paragraph{Impact of the model architecture.}
Table~\ref{tbl:evaluate-flips} shows that the vulnerable parameter ratio is mostly consistent across different DNN architectures.
However, we see that the InceptionV3 model for ImageNet has a relatively lower ratio---40.8\%---compared to the other models---between 42.1\% and 48.9\%.
We hypothesize that the reason is the auxiliary classifiers in the InceptionV3 architecture that have no function at test-time.  
To confirm our hypothesis, we simply remove the parameters in the auxiliary classifiers; which bring the vulnerability ratio closer to the other models---46.5\%.
Interestingly, we also observe that the parameters in batch normalization layers are resilient to a bit-flip: corrupting \texttt{running\_mean} and \texttt{running\_var} cause negligible damage.
In consequence, excluding the parameters in InceptionV3's multiple batch normalization layers leads to a slight increase in vulnerability---by 0.02\%.

\subsection{Implications for the Adversaries}
\label{subsec:implications-indiscriminate}
In Sec~\ref{sec:threat-model}, we defined four attack scenarios: the \MakeLowercase{\prob} and \MakeLowercase{\deter} attackers, in the black-box and white-box settings.
First, we consider the strongest attacker: the \MakeLowercase{\deter}, who can flip a bit at a specific memory location; white-box, with the model knowledge for anticipating the impact of flipping the said bit.
To carry out the attack, this attacker identifies:
\begin{enumerate*}[label=\arabic*)]
	\item how much indiscriminate damage, the RAD goal, she intends to inflict,
	\item a vulnerable parameter that can lead to the RAD goal,
	\item in this parameter, the bit location, e.g., \nth{31}-bit, and the flip direction, e.g., (0$\rightarrow$1), for inflicting the damage.
\end{enumerate*}
Based on our [RAD$>0.1$] criterion, approximately 50\% of the parameters are vulnerable in all models; thus, for this goal, the attacker can easily achieve 100\% success rate.
For more severe goals [$0.1 \leq$RAD$\leq 0.9$], our results in Appendix~\ref{append:rads-with-single-bitflips} suggest that the attacker can still find vulnerable parameters. 
In Sec~\ref{sec:rh:informed-hammer}, we discuss the necessary primitives, in a practical setting, for this attacker.

For a black-box \MakeLowercase{\deter} attacker, on the other hand, the best course of action is to target the \nth{31}-bit of a parameter.
This strategy maximizes the attacker's chance of causing indiscriminate damage, even without knowing what, or where, the corrupted parameter is.
Considering, the VGG16 model for ImageNet, the attack's success rate is 42.1\% as we report in Table~\ref{tbl:evaluate-flips}; which is an upper-bound for the black-box attackers.
For the weakest---black-box \MakeLowercase{\prob}---attacker that cannot specifically target the \nth{31}-bit, we conservatively estimate the lower-bound as 42.1\% / 32-bits = 1.32\%; assuming only the \nth{31}-bits lead to indiscriminate damage.
Note that the success rate for the white-box \MakeLowercase{\prob} attacker is still 1.32\% as acting upon the knowledge of the vulnerable parameters requires an attacker to target specific parameters.
In Sec~\ref{sec:rh:blind-hammer}, we evaluate the practical success rate of a \MakeLowercase{\prob} attacker.

\subsection{Distinct Attack Scenarios}
\label{subsec:informed-vulnerability}
In this section, other than causing indiscriminate damage, we discuss two distinct attack scenarios single bit-flips might enable: transfer learning and targeted misclassification.

\begin{figure*}[ht]
	\centering
	\includegraphics[width=0.32\textwidth]{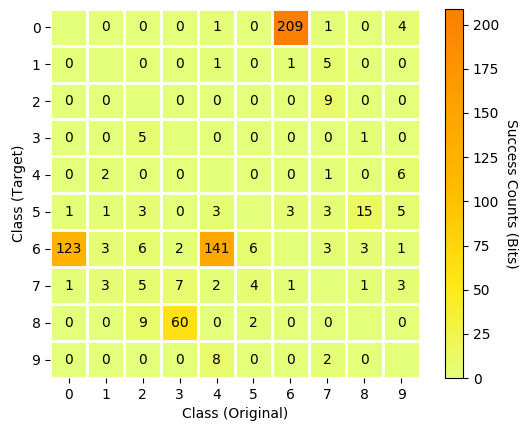}
	\includegraphics[width=0.32\textwidth]{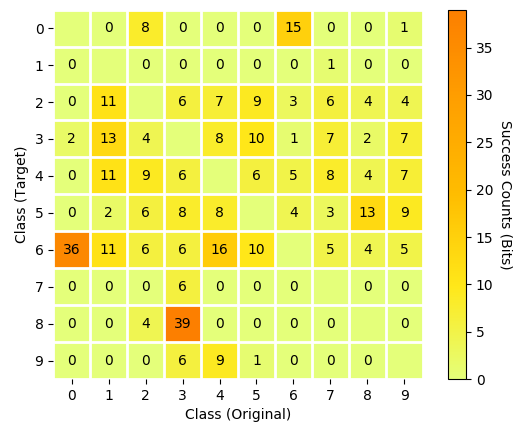}
	\includegraphics[width=0.33\textwidth]{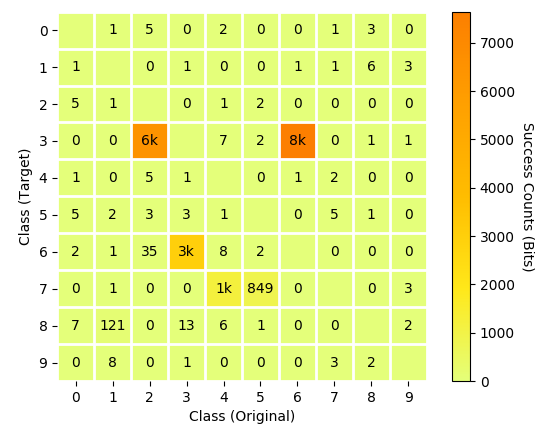}
	\caption{\textbf{
			The vulnerable parameters for a targeted attack in 3 DNN models.} Each cell reports the number of bits that lead to the misclassification of a target sample, whose original class is given by the x-axis, as the target class, which is given by the y-axis. From left to right, the models are MNIST-B, MNIST-L5 and CIFAR10-AlexNet.}
	\label{fig:tar-misclassification}
\end{figure*}

\paragraph{Transfer learning scenario.} 
Transfer learning is a common technique for \emph{transferring} the knowledge in a pre-trained \emph{teacher} model to a \emph{student} model; which, in many cases, outperforms training a model from scratch.
In a practical scenario, a service provider might rely on publicly available teacher as a starting point to train commercial student models.
The teacher's knowledge is transferred by \emph{freezing} some of its layers and embedding them into the student model; which, then, trains the remaining layers for its own task.
The security risk is that, for an attacker who knows the teacher but not the student, a black-box attack on the student might escalate into a white-box attack on the teacher's frozen layers. 
The attacker first downloads the pre-trained teacher from the Internet.
She then loads the teacher into the memory and waits for the \emph{deduplication}~\cite{xiao2013security} to happen.
During deduplication, the memory pages with the same contents---the frozen layers---are merged into the shared pages between the victim and attacker.
This essentially promotes a \MakeLowercase{\prob} threat to the victim's memory to a stronger \MakeLowercase{\deter} threat to the attacker's own memory.
In consequence, a bit-flip in the attacker's own pages can also affect the student model in the victim's memory.

We hypothesize that an attacker, who can identify the teacher's vulnerable parameters and trigger bit-flips in these parameters, can cause indiscriminate damage to the student model.
In our experiments, we examine two transfer learning tasks in~\cite{wang2018great}: the traffic sign (GTSRB)~\cite{stallkamp2012man} and flower recognition (Flower102)~\cite{nilsback2008automated}.
We initialize the student model by transferring first ten frozen layers of the teacher---VGG16 or ResNet50 on ImageNet.
We then append a new classification layer and train the resulting student network for its respective task by only updating the new unfrozen layer.
We corrupt the 1,000 parameters sampled from each layer in the teacher and monitor the damage to the student model. 
Figure~\ref{fig:deterministic} reports our results: we find that \emph{
all vulnerable parameters in the frozen layers and 
more than a half in the re-trained layers are shared by the teacher and the student}.

\paragraph{Targeted misclassification.}
Although our main focus is showing DNNs' graceless degradation, we conduct an additional experiment and ask whether a single bit-flip primitive could be used in the context of \emph{targeted misclassification} attacks.
A targeted attack aims to preserve the victim model's overall accuracy while causing it to misclassify a specific target sample into the target class.
We experiment with a target sample from each class in MNIST or CIFAR10---we use MNIST-B, MNIST-L5 and CIFAR10-AlexNet models.
Our white-box \MakeLowercase{\deter} attacker also preserves the accuracy by limiting the [RAD$<0.05$] as in~\cite{suciu2018fail}.
We find that \emph{the number of vulnerable parameters for targeted misclassifications is lower than that of for causing indiscriminate damage}. 
In Figure~\ref{fig:tar-misclassification}, we also see that for some \emph{(original--target class) pairs}, the vulnerability is more evident.
For example, in MNIST-B, there are 141 vulnerable parameters for (class $4$--class $6$) and 209 parameters for (class $6$--class $0$).
Simlarly, in CIFAR10-AlexNet, there are 6,000 parameters for (class $2$--class $3$); 3,000 parameters for (class $3$--class $6$); and 8,000 parameters for (class $6$--class $3$).
%
%

\section{Exploiting Using \rh}
\label{sec:exploit-rowhammer}

In order to corroborate the analysis made in Sec~\ref{sec:dnn-vulnerability} and prove the viability of hardware fault attacks against DNN, we test the resiliency of these models against \rh. At a high level, \rh is a software-induced fault attack that provides the attacker with a single-bit write primitive to specific physical memory locations. That is, an attacker capable of performing specific memory access patterns (at DRAM-level) can induce persistent and repeatable bit corruptions from software. Given that we focus on single-bit perturbations on DNN's parameters in practical settings, \rh represents the perfect candidate for the task. 

\begin{figure}[b]
	\definecolor{victim}{RGB}{224,118,118}
	\definecolor{aggressor}{RGB}{151,192,242}
	\centering
	\begin{minipage}[t]{0.46\linewidth}
		\captionsetup{font=footnotesize,labelfont=footnotesize}
		\includegraphics[height=2.6cm]{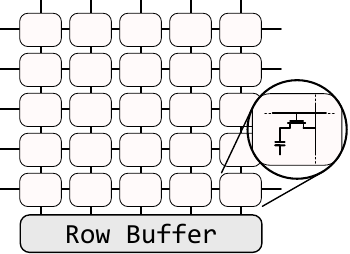}
		\caption{\textbf{DRAM bank structure.} Zoom-in on a cell containing the capacitor storing data.}
		\label{fig:dram}
	\end{minipage}
	\hspace{3mm}
	\begin{minipage}[t]{0.42\linewidth}
		\captionsetup{font=footnotesize,labelfont=footnotesize}
		\includegraphics[height=2.6cm]{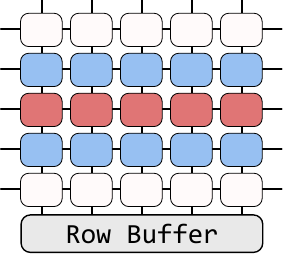}
		\caption[Double-sided \rh]{\textbf{Double-sided \rh.} Aggressor rows \tikz \draw[fill=aggressor] (0.1,0.1) rectangle (0.3,0.3);, and a victim row \tikz \draw[fill=victim] (0.1,0.1) rectangle (0.3,0.3); }.
		\label{fig:rh:ds}
	\end{minipage}
\end{figure}

\paragraph{DRAM internals.} In Figure~\ref{fig:dram}, we show the internals of a DRAM \emph{bank}. A bank is  a bi-dimensional array of memory \emph{cells} connected to a \emph{row buffer}. Every DRAM chip contains multiple banks. The cells are the actual storage of one's data. They contain a capacitor whose charge determines the value of a specific bit in memory. When a read is issued to a specific row, this row gets \emph{activated}, which means that its content gets transferred to the row buffer before being sent to the CPU. Activation is often requested to recharge a row's capacitors (i.e., \emph{refresh} operation) since they leak charge over time.

\paragraph{\rh{} mechanism.} \rh is a DRAM disturbance error that causes spurious bit-flips in DRAM cells generated by frequent activations of a neighboring row.     
Here, we focus on double-sided \rh, the most common and effective \rh variant used in practical attacks~\cite{razavi2016flip,frigo2018glitch,van2016drammer}. 
Figure~\ref{fig:rh:ds} exemplifies a typical double-sided \rh attack. 
The victim's data is stored in a row enclosed between two aggressor rows that are repeatedly accessed by the attacker. Due to the continuous activations of the neighboring rows, the victim's data is under intense duress. Thus, there is a large probability of bit-flips on its content.  

To implement such attack variant, the attacker usually needs some knowledge or control over the physical memory layout. Depending on the attack scenario, a \rh-enabled attacker can rely on a different set of primitives for this purpose. 
In our analysis, we consider two possible scenarios: 
\begin{enumerate*}[label=\arabic*)]
	\item we initially consider 
	the \emph{\MakeLowercase{\deter}} attacker; that is, an attacker with 
	the capability of causing bit-flips at the specific locations, and we 
	demonstrate how, under these assumptions, she can induce 
	indiscriminate damage to a co-located DNN application.
	\item We then deprive the attacker of this 
	ability to analyze the outcome of a \emph{\MakeLowercase{\prob}} attacker and we demonstrate that, even in a more restricted environment, 
	the attacker can still cause 
	indiscriminate damage by causing bitwise corruptions.
\end{enumerate*}

\begin{table}[t]
\centering
\adjustbox{width=\linewidth}{
	\begin{tabular}{C{1.6cm}R{2.2cm}|C{1.6cm}R{2.2cm}}
		\toprule
		\textbf{DRAM} & \multicolumn{1}{c}{\textbf{\# (0$\rightarrow$1) flips}} & \textbf{DRAM} & \multicolumn{1}{c}{\textbf{\# (0$\rightarrow$1) flips}} \\
		\midrule \midrule
		\hcell{A\_2} & \hcell{$21,538$}	& A\_4 & $5,577$ \\
		E\_2 & $16,320$ & \hcell{I\_1} & \hcell{$4,781$} \\
		H\_1 & $10,608$ & J\_1 & $4,725$ \\
		G\_1 & $7,851$  & E\_1 & $4,175$ \\
		A\_1 & $4,367$	& A\_3 & $1,541$ \\
		F\_1 & $5,927$  & \hcell{C\_1} & \hcell{$1,365$} \\
		\bottomrule
\end{tabular}
}
\caption{\textbf{Hammertime database~\cite{tatar2018defeating}.} We report the number of (0$\rightarrow$1) bit-flips in \ndimm different DRAM setups. (The rows in gray are used for the experiments in Figure~\ref{fig:timeToFlip}.)}
\label{tbl:hammertime}
\end{table}

\paragraph{Experimental setup.}
For our analysis, we constructed a simulated environment%
\footnote{We first implemented all the steps described in our paper on a physical system, considering using end-to-end attacks for our analysis. After preliminary testing of this strategy on our own DRAMs, we concluded it would be hard to generalize the findings of such an analysis and decided against it---in line with observations from prior work~\cite{tatar2018hammertime}.}
relying on a database of 
%
the \rh vulnerability in \ndimm DRAM chips,
provided by Tatar et al~\cite{tatar2018defeating}. Different memory chips have a different degree of susceptibility to the \rh vulnerability, enabling us to study the impact of \rh attacks on DNNs in different real-world scenarios. Table~\ref{tbl:hammertime} reports the susceptibility of the different memory chips to \rh. Here, we only include the numbers for (0$\rightarrow$1) bit-flips since these are the more interesting ones for the attacker targeting a DNN model according to our earlier analysis in Sec~\ref{subsec:bw_characterization} and Sec~\ref{subsec:char-model}.

We perform our analysis on an 
%
exemplary deep learning
application implemented in PyTorch, constantly querying an ImageNet model. We use ImageNet models since we focus on a scenario where the victim has a relevant memory footprint that can be realistically be targeted by hardware fault attacks such as \rh in practical settings. While small models are also potential targets, the number of interesting locations to corrupt is typically
limited to draw general conclusions on the practical effectiveness of the attack.

\subsection{\deter Attack Using \rh}
\label{sec:rh:informed-hammer}

\begin{table}[t]
\centering
\adjustbox{width=\linewidth}{
	\begin{tabular}{@{}lccc@{}}
		\toprule
		\multicolumn{1}{c}{\multirow{1}{*}{\textbf{Network}}} & \multirow{1}{*}{\textbf{Vuln. Objects}} & \multirow{1}{*}{\textbf{Vuln. Params}} & \multirow{1}{*}{\parbox{3cm}{\centering\textbf{\#Hammer Attempts}}} \\ 
		~ & \emph{(Vuln./Total)} & \emph{(in 20k params)} & \emph{(min\,/\,med\,/\,max)} \\
		\midrule \midrule
		AlexNet & $7\,/\,16$ & 9,522 & $  4\,/\,  64\,/\,4,679$\\
		VGG16 & $12\,/\,32$ & 8,140 & $  4\,/\,  64\,/\,4,679$\\
		ResNet50 & $9\,/\,102$ &  3,466 & $  4\,/\, 64\,/\,4,679$\\
		DenseNet161 & $63\,/\,806$ & 5,117 & $  4\,/\,  64\,/\,4,679$\\
		InceptionV3 & $53\,/\,483$ & 6,711 & $4\,/\,  64\,/\,4,679$\\
		\bottomrule
	\end{tabular}
}
\caption{\textbf{Effectiveness of \MakeLowercase{\deter} attacks.} We examine five different ImageNet models analyzed in Sec~\ref{sec:dnn-vulnerability}.}
\label{tbl:wb-attack}
\end{table}

We start our analysis by discussing a \MakeLowercase{\deter} attacker, 
who has the capability of causing a bit-flip at the specific location in memory. 
The two \MakeLowercase{\deter} attackers are available: the attacker with the knowledge of the victim model (white-box) and without (black-box).
%
%
However, in this section, 
we assume that the strongest attacker knows the parameters to compromise 
and is capable of triggering bit-flips on its corresponding memory location.
Then, this attacker can take advantage of accurate memory massing primitives (e.g., memory deduplication) to achieve 100\% attack success rate.



\paragraph{Memory templating.}
Since a \MakeLowercase{\deter} attacker knows the location of vulnerable parameters
, she can \emph{template} the memory up front~\cite{razavi2016flip}. That is, the attacker scans the memory by inducing \rh bit-flips in her own allocated chunks and looking for exploitable bit-flips. A \MakeLowercase{\deter} attacker aims at specific bit-flips. Hence, while templating the memory, the attacker simplifies the scan by looking for bit-flips located at specific offsets from the start address of a memory \emph{page} (i.e., 4\,KB)---the smallest possible chunk allocated from the OS. This allows the attacker to find memory pages vulnerable to \rh bit-flips at a given page offset (i.e., \emph{vulnerable templates}), which they can later use to predictably attack the victim data stored at that location.

\paragraph{Vulnerable templates.} 
To locate the parameters of the attacker's interest (i.e., vulnerable parameters) within the memory page, 
she needs to find \emph{page-aligned} data in the victim model. Modern memory allocators improve performances by storing large objects (usually multiples of the page size) page-aligned 
whereas smaller objects are not.
Thus, we first analyze the allocations performed by the PyTorch framework %
running on Python to understand if it performs such optimized page-aligned allocations for large objects similar to other programs~\cite{tcmallocAlign, jemallocAlign}. We discovered this to be the case for all the objects larger than 1\,MB---i.e., our attacker needs to 
target the parameters such as weight, bias, and so on, stored as tensor objects in layers, larger than 1\,MB.




Then, again focusing on the ImageNet models, we analyzed them to identify the  
objects that satisfy this condition. Even if the ratio between the total number of objects and target objects may seem often unbalanced in favor of the small ones%
\footnote{The \texttt{bias} in convolutional or dense layers, and the 
\texttt{running\_mean} and \texttt{running\_var} in batch-norms are usually the small objects (< 1\,MB).}, we found that the number of vulnerable parameters in the target objects is still significant (see Table~\ref{tbl:wb-attack}).
Furthermore, it is important to note that when considering a \MakeLowercase{\deter} attacker, she only needs one single vulnerable template to compromise the victim model, and there is only 1,024 possible offsets where we can store a 4-byte parameter within a 4\,KB page. 

\paragraph{Memory massaging.}
After finding a vulnerable template, the attacker needs to \emph{massage} the memory to land the victim's data on the vulnerable template. This can be achieved, for instance, by exploiting memory deduplication~\cite{razavi2016flip,bosman2016dedup,xiao2016one}. Memory deduplication is a system-level memory optimization that merges read-only pages for different processes or VMs when they contain the same data. These pages re-split when a write is issued to 
them. However, \rh behaves as invisible bit-wise writes that do not trigger the spit, breaking the process boundaries. If the attacker knows (even if partially) the content of the victim model can take advantage of this merging primitive to compromise the victim service.




\paragraph{Experimental results.} 
Based on the results of the experiments in Sec~\ref{subsec:bw_characterization} and Sec~\ref{subsec:char-model}, we analyze the requirements for a \MakeLowercase{\deter} (white-box) attacker to carry out a successful attack.
Here, we used one set of the five sampled parameters for each model.
In Table~\ref{tbl:wb-attack}, we report \emph{min, median, and max} values of the number of rows that an attacker needs to hammer to find the first vulnerable template on the \ndimm different DRAM setups for each 
model.
This provides a meaningful metric to understand the success rate of a \MakeLowercase{\deter} attack.
As you can see in Table~\ref{tbl:wb-attack}, the results remain unchanged among all the different models.
That is, for every model we tested in the best case, it required us to hammer only 4 rows (\emph{A\_2} DRAM setup) to find a vulnerable template all the way up to 4,679 in the worst case scenario (\emph{C\_1}).
The reason why the results are equal among the different models is \emph{due to the number of vulnerable parameters which largely exceeds the number of possible offsets within a page that can store such parameters} (i.e., 1024).
Since every vulnerable 
parameter yields 
indiscriminate damage [RAD$>0.1$], we simply need to identify a template that could match any given vulnerable 
parameter.
This means that an attacker can find a vulnerable template at best in a matter of few seconds\footnote{We assume 200ms to hammer a row.} and at worst still within minutes.
Once the vulnerable template is found, the attacker can leverage memory deduplication to mount an effective attack against the DNN model---with no interference with the rest of the system. 

\subsection{\prob Attack Using \rh}
\label{sec:rh:blind-hammer}

While in Sec~\ref{sec:rh:informed-hammer} we analyzed the outcome of a \MakeLowercase{\deter} attack
, here we abstract some of the assumptions made above and study the effectiveness of a \MakeLowercase{\prob} attacker oblivious of the 
bit-flip location in memory.
%
To bound the time of the lengthy \MakeLowercase{\prob} \rh attack analysis, we specifically focus our experiments on the ImageNet-VGG16 model.

We run our PyTorch application under the pressure of \rh bit-flips indiscriminately targeting both code and data regions of the process's memory.
Our goal is twofold:
\begin{enumerate*}[label=\arabic*)]
	\item to understand the effectiveness of such attack vector in a less controlled environment and
	\item to examine the robustness of a running DNN application to \rh bit-flips by measuring the number of failures (i.e., crashes) that our \MakeLowercase{\prob} attacker may inadvertently induce.
\end{enumerate*}

\paragraph{Attacker's capabilities.} 
%
We consider a \MakeLowercase{\prob} attacker who cannot control the bit-flips caused by \rh.
As a result, the attacker may corrupt bits in the DNN's parameters as well as the code blocks in the 
victim process's memory.
In principle, since \rh bit-flips propagate at the DRAM level, a fully blind \rh attacker may also inadvertently flip bits in other system memory locations.
In practice, even an attacker with limited knowledge of the system memory allocator, can heavily influence the physical memory layout by means of specially crafted memory allocations~\cite{gruss2018another,gruss2019page}.
Since this strategy allows attackers to achieve co-location with the victim memory and avoid unnecessary fault propagation in practical settings, we restrict our analysis to a scenario where bit-flips can only (blindly) corrupt memory of the victim deep learning process.
This also generalizes our analysis to arbitrary deployment scenarios, since the effectiveness of 
\MakeLowercase{\prob} attacks targeting arbitrary system memory is inherently environment-specific.

\paragraph{Methods.} For every one of the \ndimm vulnerable DRAM setups available in the database, we carried out \nRounds experiments where we performed at most \nRhSessions ``hammering'' attempts---value chosen after the \MakeLowercase{\deter} attack analysis where a median of 64 attempts was required.
The experiment has 
three possible outcomes:
\begin{enumerate*}[label=\arabic*)]
	\item we trigger one(or more) effective bit-flip(\emph{s}) that compromise the model, and we record the relative accuracy drop when performing our testing queries;
	\item we trigger one(or more) effective bit-flip(\emph{s}) in other victim memory locations that result in a crash of the deep learning process; 
	\item we reach the ``timeout'' value of \nRhSessions hammering attempts. We set such ``timeout'' value to bound our experimental analysis which would otherwise result too lengthy.
\end{enumerate*}

\begin{figure}[t]
	\centering
	\includegraphics[width=\linewidth]{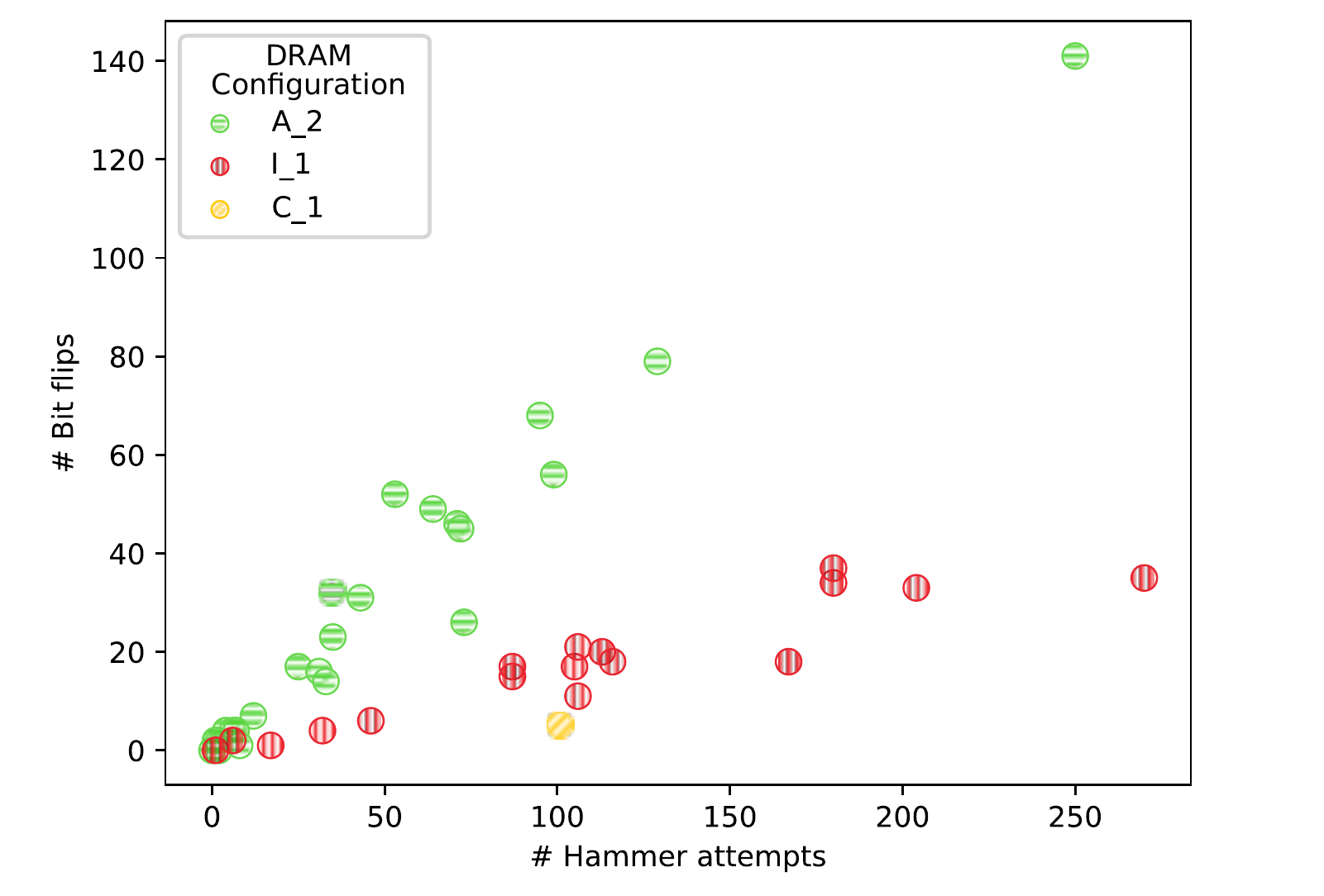}
	\vspace{-3mm}
	\caption{\textbf{The successful runs of a blind attack execution over 
	three different DRAM setups (\emph{A\_2}-most, \emph{I\_1}-least, and \emph{C\_1}-moderately vulnerable).} We report the success in terms of $\#flips$ and $\#hammer\ attempts$ required to obtain an indiscriminate damage to the victim model. 
	We observe the 
	successes within few 
	hammering attempts.}
	\label{fig:timeToFlip}
\end{figure}
\begin{figure}[t]
	\centering
	\includegraphics[width=\linewidth]{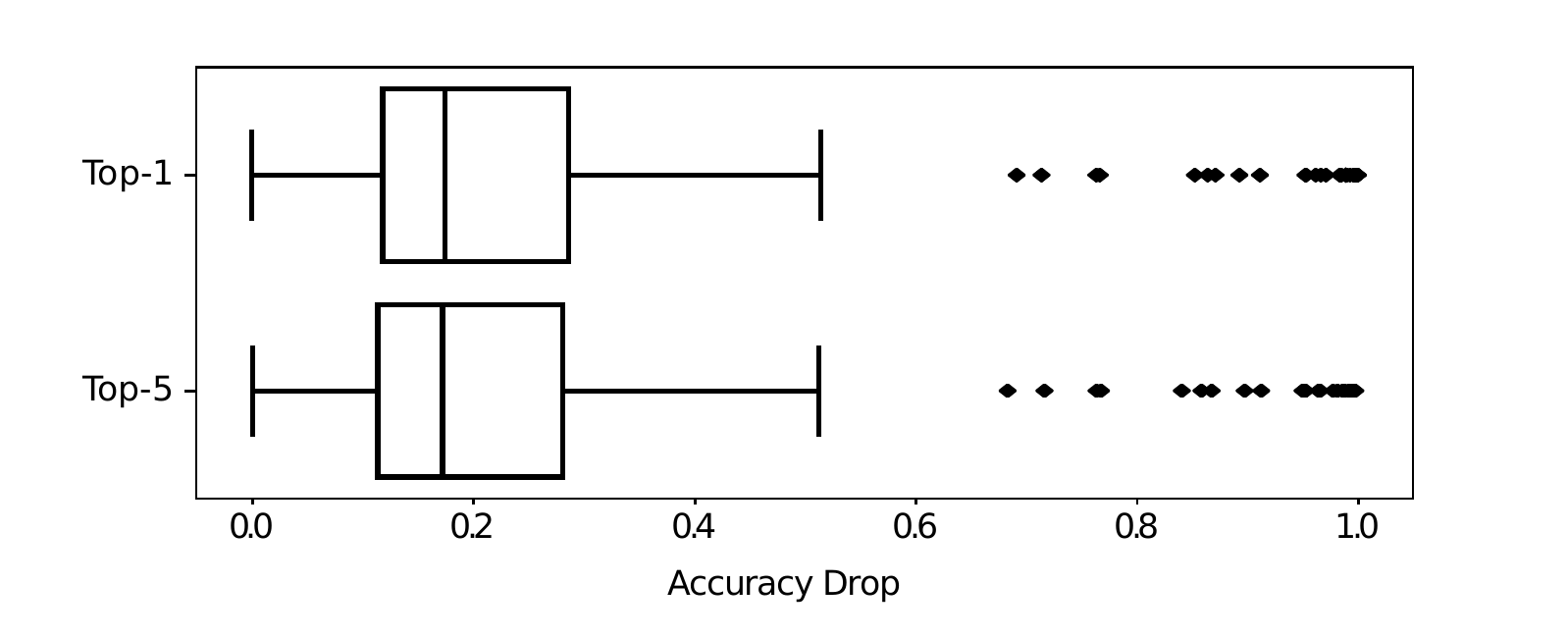}
	\vspace{-3mm}
	\caption{\textbf{The distribution of relative accuracy drop for Top-1 and Top-5.} We compute them over the effective $\#flips$ in our experiments on the ImageNet-VGG16 model.}
	\label{fig:accDrop}
\end{figure}

\paragraph{Experimental results.} In Figure~\ref{fig:timeToFlip}, we present the results for three sampled DRAM setups.
We picked \emph{A\_2}, \emph{I\_1}, and \emph{C\_1} as representative samples since they are the most, least, and moderately vulnerable DRAM chips (see Table~\ref{tbl:hammertime}).
Depending on the DRAM setup, we obtain fairly different results.
%
We found \emph{A\_2} obtains successful indiscriminate damages to the model in 24 out of \nRounds experiments while, in less vulnerable environments such as \emph{C\_1}, the number of successes decreases to only one while the other 24 times out.
However, it is important to note that a timeout does not represent a negative result
---a crash.
Contrarily, while \emph{C\_1} only had a single successful attack, it also represents a peculiar case corroborating the analysis presented in Sec~\ref{sec:dnn-vulnerability}.
The corruption generated in this single successful experiment was induced by a single bit-flip, which caused one of the most significant 
RADs detected in the entire experiment
, i.e., 0.9992 and 0.9959 in Top-1 and Top-5.
Regardless of this edge case, we report a mean of $\nDropDouble$ out of \nRounds effective attacks for this \rh variant over the different DRAM setups.
Moreover, we report the distribution of accuracy drops for Top-1 and Top-5 in Figure~\ref{fig:accDrop}.
In particular, the median drop for Top-1 and Top-5 confirms the claims made in the previous sections
, i.e., the \MakeLowercase{\prob} attacker can expect [RAD$>0.1$] on average.

Interestingly, when studying the robustness of the victim process to \rh, we discovered it to be quite resilient to spurious bit-flips.
We registered only \nCrash crashes over all the different DRAM configurations and experiments---\nRhSessions in total.
This shows that the model effectively dominates the memory footprint of the victim process and confirms findings from our earlier analysis that bit-flips in non-vulnerable model elements have essentially no noticeable impact.

\subsection{Synopsis}
\label{sec:rh:synopsis}

Throughout the section, we analyzed the outcome of \MakeLowercase{\deter} and \MakeLowercase{\prob} attacks against large DNN models and demonstrated how \rh can be deployed as a feasible attack vector against these models.
These results corroborate our findings in Sec~\ref{sec:dnn-vulnerability} where we estimated at least 40\% of a model's parameters to be vulnerable to single-bit corruptions.
Due to this large attack surface, in Sec~\ref{sec:rh:informed-hammer}, we showed that a \rh-enabled attacker armed with knowledge of the network's parameters and powerful memory massaging primitives~\cite{razavi2016flip,xiao2016one,van2016drammer} can carry out precise and effective indiscriminate attacks in a matter of, at most, few minutes in our simulated environment.
Furthermore, this property, combined with the resiliency to spurious bit-flips of the (perhaps idle) code regions, allowed us to build successful \MakeLowercase{\prob} attacks against the ImageNet-VGG16 model and inflict 
``terminal brain damage'' even when the model is hidden from the attacker.
%
%

\section{Discussion}
\label{sec:discussion}



In this section, we discuss and evaluate some potential mitigation mechanisms to protect against single-bit attacks on DNN models. 
We discuss two research directions towards making DNN models resilient to bit-flips: 
\emph{restricting activation magnitudes} and 
\emph{using low-precision numbers}.
Prior work on defenses against \rh attacks suggests system-level defenses~\cite{brasser2017can,konoth2018zebram} that often even require specific hardware support~\cite{kim2014flipping, aweke2016anvil}. 
Yet they have not been widely deployed since they require infrastructure-wide changes from cloud host providers. 
Moreover, even though the infrastructure is resilient to \rh attacks, an adversary can leverage other vectors to exploit bit-flip attacks for corrupting a model.
%
Thus, we focus on the solutions that can be directly implemented on the defended DNN model.

\subsection{Restricting Activation Magnitudes}
\label{subsec:restrict-activation}
In Sec~\ref{subsec:char-model}, we showed that the vulnerable parameter ratio varies based on inherent properties of a DNN; for instance, using PReLU activation function causes a model to propagate negative extreme activations. 
We hypothesize that an activation function, which produces its output in a constrained range, would make indiscriminate damage harder to induce via bit-flips.
There are several functions, such as Tanh or HardTanh~\cite{kalman1992tanh}, that suppress the activations; however, using ReLU-6~\cite{krizhevsky2010convolutional} function provides two key advantages over the others: 1) the defender 
only needs to substitute the existing activation functions from ReLU to ReLU6 without re-training,
and 2) ReLU$A$ functions allow the defender to control the activation range by modifying the $A$, e.g., using $A>6$ to minimize the performance loss the substitution causes.
A defender can monitor the activation values over the 
validation set and to determine an activation range that only suppresses the abnormal values, potentially caused by bit-flips. 
In our experiments on ImageNet-AlexNet, we set the 
range as $[0,max]$, where $max$ is determined adaptively by observing the maximum activation in each layer (ReLU-A).


\begin{table}[t]
\centering
\adjustbox{width=\linewidth}{
	\begin{tabular}{@{}lcccc@{}}
		\toprule		
		\multicolumn{1}{c}{\textbf{Network}} & \textbf{Train} & \textbf{Base acc.} & \textbf{\# Params} & \textbf{Vulnerability} \\ \midrule \midrule
		Base (ReLU) & Scr & 98.13 & \multirow{4}{*}{21,840} & 10,972 (50.2\%) \\
		Base (ReLU6) & Scr & 98.16 &  & 313 (1.4\%) \\
		Base (Tanh) & Scr & 97.25 &  & 507 (2.3\%) \\
		Base (ReLU6) & Sub & 95.71 &  & 542 (2.4\%) \\ \midrule \midrule
		AlexNet (ReLU) & - & 56.52 / 79.07 & \multirow{3}{*}{\begin{tabular}[c]{@{}c@{}}20,000 \\ (61M)\end{tabular}} & 9.467 (47.34\%) \\
		AlexNet (ReLU6) & Sub & 39.80 / 65.82 &  & 560 (2.8\%) \\
		AlexNet (ReLUA) & Sub & 56.52 / 79.07 &  & 1,063 (5.32\%) \\ \midrule \midrule
		VGG16 (ReLU) & - & 64.28 / 86.56 & \multirow{3}{*}{\begin{tabular}[c]{@{}c@{}}20,000 \\ (138M)\end{tabular}} & 8,227 (41.13\%) \\
		VGG16 (ReLU6) & Sub & 38.58 / 64.84 &  & 2,339 (11.67\%) \\
		VGG16 (ReLUA) & Sub & 64.28 / 86.56 &  & 2,427 (12.14\%) \\ \bottomrule
	\end{tabular}
}
\caption{\textbf{Effectiveness of restricting activation.}}
\label{tbl:defense-suppress}
\end{table}

\paragraph{Experiments.} We use three DNN models in Sec~\ref{sec:dnn-vulnerability}: the MNIST-B, 
ImageNet-AlexNet, and ImageNet-VGG16 models. 
We evaluate four activation functions: ReLU (default), Tanh, ReLU6, and ReLU$A$ (only for AlexNet and VGG16); and two training methods: training a model from scratch (\emph{Scr}) or substituting the existing activation into another (\emph{Sub}). 
In our notation, we denote the model's name together with its activation function, e.g., AlexNet (ReLU6)
For larger models, we also rely on our speed-up heuristics for estimating the vulnerability.

%
Table~\ref{tbl:defense-suppress} shows the effectiveness of this defensive mechanism. 
For each network (Column 1), we list the training method, the base accuracy, the number of sampled parameters, and the vulnerability (Column 2-5). 
We find that, in some cases, restricting activation magnitudes with Tanh and ReLU6 reduces the vulnerability. 
For instance, in the MNIST models, we observe that the ratio of vulnerable parameters drops to 1.4-2.4\% from 50\%; without incurring any significant performance loss.
Further, we discover that the substitution with ReLU6 achieves a similar effect without re-training; however, it fails to prevent the vulnerability in the last layer, which uses Softmax instead of ReLU.
In AlexNet and VGG16, we also observe a decrease in the ratio of vulnerable parameters---47.34\% to 2.8\% and 41.13\% to 11.67\%---; however, with significant loss of accuracy. 
To minimize the loss, we set the range of activation in AlexNet (ReLU$A$) and VGG16 (ReLU$A$) by selecting the maximum activation value in each layer. 
We see that ReLU$A$ leads to a trade-off between the ratio of vulnerable parameters and the accuracy.

\paragraph{Takeaways.} Our experimental results on restricting activation magnitudes suggest that this mechanism 1) allows a defender to control the trade-off between the relative accuracy drop and reducing the vulnerable parameters and 2) enables ad-hoc defenses to DNN models, which does not require training the network from scratch. However, the remaining number of vulnerable parameters shows that the \rh attacker still could inflict damage, with a reduced success rate.

\subsection{Using Low-precision Numbers}
\label{subsec:low-precisions}


\begin{table}[t]
	\centering
	\adjustbox{width=\linewidth}{
		\begin{tabular}{@{}ccccc@{}}
			\toprule
			\textbf{Network} & \textbf{Method} & \textbf{Base acc.} & \textbf{\# Params} & \textbf{Vulnerability} \\ \midrule \midrule
			L5 & - & 99.24 & 62,598 & 30,686 (49.0\%) \\
			L5 & 8-bit Quantized & 99.03 & 62,600 & 0 (0.0\%) \\
			L5 & XNOR Binarized & 98.39 & 62,286 & 623 (1.0\%) \\ \bottomrule
		\end{tabular}
	}
	\caption{\textbf{Effectiveness of using low-precision.}}
	\label{tbl:defense-lprecision}
\end{table}

Another direction is to represent the model parameters as low-precision numbers by using quantization and binarization. 
In Sec~\ref{subsec:bw_characterization}, we found that the vulnerability exploits the bitwise representation of the corrupted parameter to induce the dramatic chances in the parameter's value. 
We hypothesize that the use of low-precision numbers would make a parameter more resilient to such changes. 
For example, an integer expressed as the 8-bit quantized format can be increased at most 128 by a flip in the most significant bit---\nth{8} bit. 
Therefore, the attacker only can cause restricted increases in a model parameter. 
Training models using low-precision numbers are supported by the popular deep learning frameworks such as TensorFlow\footnote{\url{https://www.tensorflow.org/lite/performance/post\_training\_quantization}}. 
The victim can train and deploy the model with quantized or binarized parameters on these frameworks.

\paragraph{Experiments.} 
To test our hypothesis, we use 3 DNN models: the MNIST-L5 (baseline) and its quantized and binarized variants.
To quantize the MNIST-L5 model, we use the 8-bit quantization in~\cite{wang2018training,banner2018scalable}, which converts the model parameters in all layers into integers between 0 and 255. 
For the binarization, we employ XNOR-Net~\cite{rastegari2016xnor}, which converts the model parameters to -1 and 1, except for the first convolutional layer. 
Using these variants, we evaluate the vulnerability to single bit-flips and report the accuracy, total parameters and vulnerability; without the speed-up heuristics.

Table~\ref{tbl:defense-lprecision} shows the effectiveness of using low-precision parameters. 
For each network (Column 1), we report the quantization method, the accuracy, the number of vulnerable parameters and their percentage (Columns 2-5). 
We find that \emph{using low-precision parameters reduces the vulnerability}: in all cases, the ratio of vulnerable parameters drops from 49\% (Baseline) to 0-2\% (surprisingly 0\% with the quantization). 
We also observe that, in the binarized model, the first convolutional and the last classification layers contain most of the vulnerable parameters; with 150 and 420 vulnerable parameters, respectively.
This also corroborates with our results in Sec~\ref{subsec:bw_characterization}.

\paragraph{Takeaways.} 
Even though 8-bit quantization mitigates the vulnerability, in a real-world scenario, training a large model, such as~\cite{wu2016google}, from scratch can take a week on a supercomputing cluster.
This computational burden lessens the practicality of this defensive mechanism.

%

%
%

\section{Related Work}
\label{sec:related}




\paragraph{DNN's resilience to perturbations.}
Prior work has utilized the \emph{graceful degredation} of DNN models under parameter perturbations in a wide range of applications. 
For example, network quantization~\cite{anwar2015fixed,zhou2017incremental}, by quantizing a DNN model's high-precision parameter into low-precision, reduces the size and 
inference time of a model with 
negligible performance penalty. 
This property has also been used as a primitive for improving the security of DNNs. 
For example, modifying the parameter slightly to inject a watermark to allow model owners to prove ownership~\cite{adi2018turning}; adding Gaussian noise to model parameter 
for reducing the reliability of test-time adversarial attacks on DNNs~\cite{zhou2018breaking}; and fine-tuning the parameters for mitigating the malicious backdoors in a model~\cite{liu2018fine}. 
Further, the resilience to structural changes has lead to pruning techniques~\cite{han2015deep,li2016pruning,anwar2017structured} which improve the efficiency of a DNN model by removing unimportant neurons along with their parameters.
In our work, we study the \emph{graceless degredation} of DNNs under hardware fault attacks that induce single bit-flips in individual parameters.

\paragraph{Indiscriminate poisoning attacks on DNNs.} 
Recent work on adversarial machine learning has demonstrated many attack scenarios to inflict indiscriminate damage on a model. 
One of the well-studied vectors is \emph{indiscriminate poisoning attacks}~\cite{Biggio:2012:PAA:3042573.3042761} in which the adversary, by injecting malicious data in the victim's training set, aims to hurt the model.
Previous studies suggest that such attack might require significant amount of poisonous instances~\cite{nelson2008exploiting}. 
For example, Steinhardt et al.~\cite{steinhardt2017certified} shows that, with IMDB dataset, an attacker needs to craft 3\% of the total training instances to achieve 11\% of accuracy drop compared to the pristine model. 
Further, the defenses based on robust outlier removal techniques could render poison injection ineffective by filtering it out~\cite{steinhardt2017certified, diakonikolas2018sever}. 
Moreover, to achieve targeted damages without harming the model's overall accuracy, \emph{targeted poisoning attacks}~\cite{suciu2018fail, shafahi2018poison} have been studied. 
In this paper, we analyze a test-time vulnerability that does not require the adversary's contact to the victim model during its training. 
This vulnerability inflicts indiscriminate damage, similar to indiscriminate poisoning attacks, through a different attack medium.

\paragraph{Hardware fault injection attacks.}
Hardware fault injection is a class of attacks that rely on hardware glitches on the system to corrupt victim's data. 
These glitches generally provide a single-bit write primitive at the physical memory; which could potentially lead to privilege escalation~\cite{xiao2016one}.
While in the past these attacks required physical access to the victim's system~\cite{breier2018practical,liu2017fault}, recently they have gained more momentum since the software-based version of these attacks were demonstrated~\cite{kim2014flipping,tang2017clkscrew}. 
Instances of these attacks are 1) the \clkscrew attack~\cite{tang2017clkscrew} that leverages dynamic voltage and frequency scaling on mobile processors generate faults on instructions; or 2) the well-known \rh vulnerability that triggers bitwise corruptions in DRAM. 
\rh has been used in the context of cloud VMs~\cite{razavi2016flip, xiao2016one}, on desktops~\cite{seaborn2015rh} and mobile~\cite{van2016drammer} and even to compromise browsers from JavaScript~\cite{gruss2016rhjs, bosman2016dedup, frigo2018glitch}. 
In the context of DNNs, fault attacks have been proposed as an alternative for inflicting indiscriminate damages.
Instead of injecting poisonous instances, fault attacks directly induce perturbations to the models running on hardware~\cite{liu2017fault, li2017understanding, breier2018practical, reagen2018ares, clements2018hardware}. 
These studies have considered the adversaries with direct access to the victim hardware~\cite{clements2018hardware,breier2018practical} and adversaries who randomly corrupt parameters~\cite{liu2017fault,reagen2018ares,li2017understanding}.
We utilize \rh as an established fault attack to demonstrate practical implications of the graceless degradation of DNNs.
Our threat model follows the realistic single bit-flip capability of a fault attack and modern application of DNNs in a cloud environment, where physical access to the hardware is impractical.
%
%

\section{Conclusions}
\label{sec:conclusions}
This work exposes the limits of DNN's resilience against the parameter perturbations. We study the vulnerability of DNN models to single bit-flips.
We evaluate 19 DNN models with six architectures on three image classification tasks and estimate that 40-50\% of a DNN's parameters are vulnerable. 
An attacker, with only a single-bit corruption of these vulnerable parameters, can cause indiscriminate damage [RAD$>0.1$].
We further characterize this vulnerability based on the impact of various factors: the bit position, bit-flip direction, parameter sign, layer width, activation function, training techniques, and model architecture. 
To demonstrate the feasibility of the bit-flip attacks in practice, we leverage a software-induced fault injection attack, \rh.
In experiments with \rh, we find that, without knowing the victim's deep learning system, the attacker can inflict indiscriminate damage without system crashes.
Lastly, motivated by the attacks, we discuss two potential directions of mitigation: restricting activation magnitudes and using low-precision numbers.
We believe that our work is an important step for understanding and mitigating this emerging threat that can compromise the security of critical deep learning systems.

\section*{Acknowledgments}

We thank Tom Goldstein, Dana Dachman-Soled, our shepherd, David Evans, and the anonymous reviewers for their feedback.
We also acknowledge the University of Maryland super-computing resources\footnote{\url{http://hpcc.umd.edu}} (DeepThought2) made available for conducting the experiments reported in our paper. 
This research was partially supported by the Department of Defense, by the United States Office of Naval Research (ONR) under contract N00014-17-1-2782 (BinRec), by the European Union's Horizon 2020 research and innovation programme under grant agreement No. 786669 (ReAct) and No. 825377 (UNICORE), and by the Netherlands Organisation for Scientific Research through grant NWO 639.021.753 VENI (PantaRhei). This paper reflects only the authors' view. The funding agencies are not responsible for any use that may be made of the information it contains.



{
	\footnotesize
	\bibliographystyle{plain}
	
	\bibliography{bibliography/security,bibliography/additionals}

\begin{thebibliography}{10}

\bibitem{adi2018turning}
Yossi Adi, Carsten Baum, Moustapha Cisse, Benny Pinkas, and Joseph Keshet.
\newblock Turning your weakness into a strength: Watermarking deep neural
  networks by backdooring.
\newblock In {\em 27th {USENIX} Security Symposium ({USENIX} Security 18)},
  pages 1615--1631, Baltimore, MD, 2018. {USENIX} Association.

\bibitem{an1996effects}
G.~{An}.
\newblock The effects of adding noise during backpropagation training on a
  generalization performance.
\newblock {\em Neural Computation}, 8(3):643--674, April 1996.

\bibitem{anwar2015fixed}
Sajid Anwar, Kyuyeon Hwang, and Wonyong Sung.
\newblock Fixed point optimization of deep convolutional neural networks for
  object recognition.
\newblock In {\em Acoustics, Speech and Signal Processing (ICASSP), 2015 IEEE
  International Conference on}, pages 1131--1135. IEEE, 2015.

\bibitem{anwar2017structured}
Sajid Anwar, Kyuyeon Hwang, and Wonyong Sung.
\newblock Structured pruning of deep convolutional neural networks.
\newblock {\em ACM Journal on Emerging Technologies in Computing Systems
  (JETC)}, 13(3):32, 2017.

\bibitem{zhou2017incremental}
Yiwen Guo Lin Xu Yurong~Chen Aojun~Zhou, Anbang~Yao.
\newblock Incremental network quantization: Towards lossless cnns with
  low-precision weights.
\newblock In {\em International Conference on Learning Representations
  ({ICLR})}, 2017.

\bibitem{aweke2016anvil}
Zelalem~Birhanu Aweke, Salessawi~Ferede Yitbarek, Rui Qiao, Reetuparna Das,
  Matthew Hicks, Yossi Oren, and Todd Austin.
\newblock Anvil: Software-based protection against next-generation rowhammer
  attacks.
\newblock {\em ACM SIGPLAN Notices}, 51(4):743--755, 2016.

\bibitem{banner2018scalable}
Ron Banner, Itay Hubara, Elad Hoffer, and Daniel Soudry.
\newblock Scalable methods for 8-bit training of neural networks.
\newblock In S.~Bengio, H.~Wallach, H.~Larochelle, K.~Grauman, N.~Cesa-Bianchi,
  and R.~Garnett, editors, {\em Advances in Neural Information Processing
  Systems 31}, pages 5145--5153. Curran Associates, Inc., 2018.

\bibitem{Biggio:2012:PAA:3042573.3042761}
Battista Biggio, Blaine Nelson, and Pavel Laskov.
\newblock Poisoning attacks against support vector machines.
\newblock In {\em Proceedings of the 29th International Coference on
  International Conference on Machine Learning}, ICML'12, pages 1467--1474,
  USA, 2012. Omnipress.

\bibitem{bosman2016dedup}
Erik Bosman, Kaveh Razavi, Herbert Bos, and Cristiano Giuffrida.
\newblock Dedup est machina: Memory deduplication as an advanced exploitation
  vector.
\newblock In {\em 2016 IEEE symposium on security and privacy (SP)}, pages
  987--1004. IEEE, 2016.

\bibitem{brasser2017can}
Ferdinand Brasser, Lucas Davi, David Gens, Christopher Liebchen, and Ahmad-Reza
  Sadeghi.
\newblock Can{\textquoteright}t touch this: Software-only mitigation against
  rowhammer attacks targeting kernel memory.
\newblock In {\em 26th {USENIX} Security Symposium ({USENIX} Security 17)},
  pages 117--130, Vancouver, BC, 2017. {USENIX} Association.

\bibitem{breier2018practical}
Jakub Breier, Xiaolu Hou, Dirmanto Jap, Lei Ma, Shivam Bhasin, and Yang Liu.
\newblock Practical fault attack on deep neural networks.
\newblock In {\em Proceedings of the 2018 ACM SIGSAC Conference on Computer and
  Communications Security}, CCS '18, pages 2204--2206, New York, NY, USA, 2018.
  ACM.

\bibitem{chen2015deepdriving}
Chenyi Chen, Ari Seff, Alain Kornhauser, and Jianxiong Xiao.
\newblock Deepdriving: Learning affordance for direct perception in autonomous
  driving.
\newblock In {\em Proceedings of the IEEE International Conference on Computer
  Vision}, pages 2722--2730, 2015.

\bibitem{clements2018hardware}
Joseph Clements and Yingjie Lao.
\newblock Hardware trojan attacks on neural networks, 2018.

\bibitem{diakonikolas2018sever}
Ilias Diakonikolas, Gautam Kamath, Daniel~M. Kane, Jerry Li, Jacob Steinhardt,
  and Alistair Stewart.
\newblock Sever: A robust meta-algorithm for stochastic optimization, 2018.

\bibitem{frigo2018glitch}
Pietro Frigo, Cristiano Giuffrida, Herbert Bos, and Kaveh Razavi.
\newblock Grand pwning unit: accelerating microarchitectural attacks with the
  gpu.
\newblock In {\em Grand Pwning Unit: Accelerating Microarchitectural Attacks
  with the GPU}, page~0. IEEE, 2018.

\bibitem{tcmallocAlign}
Sanjay Ghemawat.
\newblock Tcmalloc : Thread-caching malloc, 2018.

\bibitem{gruss2019page}
Daniel Gruss, Erik Kraft, Trishita Tiwari, Michael Schwarz, Ari Trachtenberg,
  Jason Hennessey, Alex Ionescu, and Anders Fogh.
\newblock Page cache attacks.
\newblock {\em arXiv preprint arXiv:1901.01161}, 2019.

\bibitem{gruss2018another}
Daniel Gruss, Moritz Lipp, Michael Schwarz, Daniel Genkin, Jonas Juffinger,
  Sioli O'Connell, Wolfgang Schoechl, and Yuval Yarom.
\newblock Another flip in the wall of rowhammer defenses.
\newblock In {\em 2018 IEEE Symposium on Security and Privacy (SP)}, pages
  245--261. IEEE, 2018.

\bibitem{gruss2016rhjs}
Daniel Gruss, Cl{\'e}mentine Maurice, and Stefan Mangard.
\newblock Rowhammer. js: A remote software-induced fault attack in javascript.
\newblock In {\em International Conference on Detection of Intrusions and
  Malware, and Vulnerability Assessment}, pages 300--321. Springer, 2016.

\bibitem{han2015deep}
Song Han, Huizi Mao, and William~J Dally.
\newblock Deep compression: Compressing deep neural networks with pruning,
  trained quantization and huffman coding.
\newblock {\em International Conference on Learning Representations (ICLR)},
  2016.

\bibitem{he2015delving}
Kaiming He, Xiangyu Zhang, Shaoqing Ren, and Jian Sun.
\newblock Delving deep into rectifiers: Surpassing human-level performance on
  imagenet classification.
\newblock In {\em Proceedings of the IEEE international conference on computer
  vision}, pages 1026--1034, 2015.

\bibitem{he2016deep}
Kaiming He, Xiangyu Zhang, Shaoqing Ren, and Jian Sun.
\newblock Deep residual learning for image recognition.
\newblock In {\em Proceedings of the IEEE conference on computer vision and
  pattern recognition}, pages 770--778, 2016.

\bibitem{iandola2014densenet}
Forrest Iandola, Matt Moskewicz, Sergey Karayev, Ross Girshick, Trevor Darrell,
  and Kurt Keutzer.
\newblock Densenet: Implementing efficient convnet descriptor pyramids.
\newblock {\em arXiv preprint arXiv:1404.1869}, 2014.

\bibitem{ioffe2015batch}
Sergey Ioffe and Christian Szegedy.
\newblock Batch normalization: Accelerating deep network training by reducing
  internal covariate shift.
\newblock In Francis Bach and David Blei, editors, {\em Proceedings of the 32nd
  International Conference on Machine Learning}, volume~37 of {\em Proceedings
  of Machine Learning Research}, pages 448--456, Lille, France, 07--09 Jul
  2015. PMLR.

\bibitem{kalman1992tanh}
Barry~L Kalman and Stan~C Kwasny.
\newblock Why tanh: Choosing a sigmoidal function.
\newblock In {\em Neural Networks, 1992. IJCNN., International Joint Conference
  on}, volume~4, pages 578--581. IEEE, 1992.

\bibitem{kim2014flipping}
Yoongu Kim, Ross Daly, Jeremie Kim, Chris Fallin, Ji~Hye Lee, Donghyuk Lee,
  Chris Wilkerson, Konrad Lai, and Onur Mutlu.
\newblock Flipping bits in memory without accessing them: An experimental study
  of dram disturbance errors.
\newblock In {\em ACM SIGARCH Computer Architecture News}, volume~42, pages
  361--372. IEEE Press, 2014.

\bibitem{konoth2018zebram}
Radhesh~Krishnan Konoth, Marco Oliverio, Andrei Tatar, Dennis Andriesse,
  Herbert Bos, Cristiano Giuffrida, and Kaveh Razavi.
\newblock Zebram: Comprehensive and compatible software protection against
  rowhammer attacks.
\newblock In {\em 13th {USENIX} Symposium on Operating Systems Design and
  Implementation ({OSDI} 18)}, pages 697--710, Carlsbad, CA, 2018. {USENIX}
  Association.

\bibitem{krizhevsky2010convolutional}
Alex Krizhevsky and Geoff Hinton.
\newblock Convolutional deep belief networks on cifar-10.
\newblock {\em Unpublished manuscript}, 40(7), 2010.

\bibitem{krizhevsky2009learning}
Alex Krizhevsky and Geoffrey Hinton.
\newblock Learning multiple layers of features from tiny images.
\newblock Technical report, Citeseer, 2009.

\bibitem{krizhevsky2012imagenet}
Alex Krizhevsky, Ilya Sutskever, and Geoffrey~E Hinton.
\newblock Imagenet classification with deep convolutional neural networks.
\newblock In {\em Advances in neural information processing systems}, pages
  1097--1105, 2012.

\bibitem{lecun1998gradient}
Yann LeCun, L{\'e}on Bottou, Yoshua Bengio, and Patrick Haffner.
\newblock Gradient-based learning applied to document recognition.
\newblock {\em Proceedings of the IEEE}, 86(11):2278--2324, 1998.

\bibitem{lecun1990optimal}
Yann LeCun, John~S Denker, and Sara~A Solla.
\newblock Optimal brain damage.
\newblock In {\em Advances in neural information processing systems}, pages
  598--605, 1990.

\bibitem{lecun2015lenet}
Yann LeCun et~al.
\newblock Lenet-5, convolutional neural networks.
\newblock {\em URL: http://yann. lecun. com/exdb/lenet}, page~20, 2015.

\bibitem{li2017understanding}
Guanpeng Li, Siva Kumar~Sastry Hari, Michael Sullivan, Timothy Tsai, Karthik
  Pattabiraman, Joel Emer, and Stephen~W Keckler.
\newblock Understanding error propagation in deep learning neural network (dnn)
  accelerators and applications.
\newblock In {\em Proceedings of the International Conference for High
  Performance Computing, Networking, Storage and Analysis}, page~8. ACM, 2017.

\bibitem{li2016pruning}
Hao Li, Asim Kadav, Igor Durdanovic, Hanan Samet, and Hans~Peter Graf.
\newblock Pruning filters for efficient convnets.
\newblock {\em arXiv preprint arXiv:1608.08710}, 2016.

\bibitem{lipp2018nethammer}
Moritz Lipp, Misiker~Tadesse Aga, Michael Schwarz, Daniel Gruss, Cl{\'e}mentine
  Maurice, Lukas Raab, and Lukas Lamster.
\newblock Nethammer: Inducing rowhammer faults through network requests.
\newblock {\em arXiv preprint arXiv:1805.04956}, 2018.

\bibitem{liu2018fine}
Kang Liu, Brendan Dolan-Gavitt, and Siddharth Garg.
\newblock Fine-pruning: Defending against backdooring attacks on deep neural
  networks.
\newblock In {\em Research in Attacks, Intrusions, and Defenses ({RAID})},
  pages 273--294, 2018.

\bibitem{liu2017fault}
Yannan Liu, Lingxiao Wei, Bo~Luo, and Qiang Xu.
\newblock Fault injection attack on deep neural network.
\newblock In {\em Proceedings of the 36th International Conference on
  Computer-Aided Design}, pages 131--138. IEEE Press, 2017.

\bibitem{jemallocAlign}
Jemalloc manual.
\newblock {Jemalloc: general purpose memory allocation functions}.
\newblock \url{http://jemalloc.net/jemalloc.3.html}, 2019.

\bibitem{nelson2008exploiting}
Blaine Nelson, Marco Barreno, Fuching~Jack Chi, Anthony~D Joseph, Benjamin~IP
  Rubinstein, Udam Saini, Charles~A Sutton, J~Doug Tygar, and Kai Xia.
\newblock Exploiting machine learning to subvert your spam filter.
\newblock {\em LEET}, 8:1--9, 2008.

\bibitem{nilsback2008automated}
Maria-Elena Nilsback and Andrew Zisserman.
\newblock Automated flower classification over a large number of classes.
\newblock In {\em Computer Vision, Graphics \& Image Processing, 2008.
  ICVGIP'08. Sixth Indian Conference on}, pages 722--729. IEEE, 2008.

\bibitem{qin2017robustness}
Minghai Qin, Chao Sun, and Dejan Vucinic.
\newblock Robustness of neural networks against storage media errors, 2017.

\bibitem{rastegari2016xnor}
Mohammad Rastegari, Vicente Ordonez, Joseph Redmon, and Ali Farhadi.
\newblock Xnor-net: Imagenet classification using binary convolutional neural
  networks.
\newblock In {\em ECCV}, 2016.

\bibitem{razavi2016flip}
Kaveh Razavi, Ben Gras, Erik Bosman, Bart Preneel, Cristiano Giuffrida, and
  Herbert Bos.
\newblock Flip feng shui: Hammering a needle in the software stack.
\newblock In {\em 25th {USENIX} Security Symposium ({USENIX} Security 16)},
  pages 1--18, Austin, TX, 2016. {USENIX} Association.

\bibitem{reagen2018ares}
Brandon Reagen, Udit Gupta, Lillian Pentecost, Paul Whatmough, Sae~Kyu Lee,
  Niamh Mulholland, David Brooks, and Gu-Yeon Wei.
\newblock Ares: A framework for quantifying the resilience of deep neural
  networks.
\newblock In {\em 2018 55th ACM/ESDA/IEEE Design Automation Conference (DAC)},
  pages 1--6. IEEE, 2018.

\bibitem{ilsvrc2012}
Olga Russakovsky, Jia Deng, Hao Su, Jonathan Krause, Sanjeev Satheesh, Sean Ma,
  Zhiheng Huang, Andrej Karpathy, Aditya Khosla, Michael Bernstein,
  Alexander~C. Berg, and Li~Fei-Fei.
\newblock {ImageNet Large Scale Visual Recognition Challenge}.
\newblock {\em International Journal of Computer Vision (IJCV)},
  115(3):211--252, 2015.

\bibitem{russakovsky2015imagenet}
Olga Russakovsky, Jia Deng, Hao Su, Jonathan Krause, Sanjeev Satheesh, Sean Ma,
  Zhiheng Huang, Andrej Karpathy, Aditya Khosla, Michael Bernstein, et~al.
\newblock Imagenet large scale visual recognition challenge.
\newblock {\em International Journal of Computer Vision}, 115(3):211--252,
  2015.

\bibitem{seaborn2015rh}
Mark Seaborn and Thomas Dullien.
\newblock Exploiting the dram rowhammer bug to gain kernel privileges.

\bibitem{shafahi2018poison}
Ali Shafahi, W.~Ronny Huang, Mahyar Najibi, Octavian Suciu, Christoph Studer,
  Tudor Dumitras, and Tom Goldstein.
\newblock Poison frogs! targeted clean-label poisoning attacks on neural
  networks.
\newblock In S.~Bengio, H.~Wallach, H.~Larochelle, K.~Grauman, N.~Cesa-Bianchi,
  and R.~Garnett, editors, {\em Advances in Neural Information Processing
  Systems 31}, pages 6106--6116. Curran Associates, Inc., 2018.

\bibitem{simonyan2014very}
Karen Simonyan and Andrew Zisserman.
\newblock Very deep convolutional networks for large-scale image recognition.
\newblock {\em International Conference on Learning Representations (ICLR)},
  2015.

\bibitem{smolyanskiy2017toward}
Nikolai Smolyanskiy, Alexey Kamenev, Jeffrey Smith, and Stan Birchfield.
\newblock Toward low-flying autonomous mav trail navigation using deep neural
  networks for environmental awareness.
\newblock In {\em 2017 IEEE/RSJ International Conference on Intelligent Robots
  and Systems (IROS)}, pages 4241--4247. IEEE, 2017.

\bibitem{srivastava2014dropout}
Nitish Srivastava, Geoffrey Hinton, Alex Krizhevsky, Ilya Sutskever, and Ruslan
  Salakhutdinov.
\newblock Dropout: A simple way to prevent neural networks from overfitting.
\newblock {\em The Journal of Machine Learning Research}, 15(1):1929--1958,
  2014.

\bibitem{stallkamp2012man}
Johannes Stallkamp, Marc Schlipsing, Jan Salmen, and Christian Igel.
\newblock Man vs. computer: Benchmarking machine learning algorithms for
  traffic sign recognition.
\newblock {\em Neural networks}, 32:323--332, 2012.

\bibitem{steinhardt2017certified}
Jacob Steinhardt, Pang Wei~W Koh, and Percy~S Liang.
\newblock Certified defenses for data poisoning attacks.
\newblock In {\em Advances in Neural Information Processing Systems}, pages
  3517--3529, 2017.

\bibitem{suciu2018fail}
Octavian Suciu, Radu Marginean, Yigitcan Kaya, Hal~Daume III, and Tudor
  Dumitras.
\newblock When does machine learning {FAIL}? generalized transferability for
  evasion and poisoning attacks.
\newblock In {\em 27th {USENIX} Security Symposium ({USENIX} Security 18)},
  pages 1299--1316, Baltimore, MD, 2018. {USENIX} Association.

\bibitem{szegedy2016rethinking}
Christian Szegedy, Vincent Vanhoucke, Sergey Ioffe, Jon Shlens, and Zbigniew
  Wojna.
\newblock Rethinking the inception architecture for computer vision.
\newblock In {\em Proceedings of the IEEE conference on computer vision and
  pattern recognition}, pages 2818--2826, 2016.

\bibitem{tang2017clkscrew}
Adrian Tang, Simha Sethumadhavan, and Salvatore Stolfo.
\newblock {CLKSCREW}: Exposing the perils of security-oblivious energy
  management.
\newblock In {\em 26th {USENIX} Security Symposium ({USENIX} Security 17)},
  pages 1057--1074, Vancouver, BC, 2017. {USENIX} Association.

\bibitem{tatar2018defeating}
Andrei Tatar, Cristiano Giuffrida, Herbert Bos, and Kaveh Razavi.
\newblock Defeating software mitigations against rowhammer: a surgical
  precision hammer.
\newblock In {\em International Symposium on Research in Attacks, Intrusions,
  and Defenses}, pages 47--66. Springer, 2018.

\bibitem{tatar2018hammertime}
Andrei Tatar, Cristiano Giuffrida, Herbert Bos, and Kaveh Razavi.
\newblock Defeating software mitigations against rowhammer: a surgical
  precision hammer.
\newblock In {\em International Symposium on Research in Attacks, Intrusions,
  and Defenses}, pages 47--66. Springer, 2018.

\bibitem{tatar2018throwhammer}
Andrei Tatar, Radhesh~Krishnan Konoth, Elias Athanasopoulos, Cristiano
  Giuffrida, Herbert Bos, and Kaveh Razavi.
\newblock Throwhammer: Rowhammer attacks over the network and defenses.
\newblock In {\em 2018 {USENIX} Annual Technical Conference ({USENIX} {ATC}
  18)}, pages 213--226, Boston, MA, 2018. {USENIX} Association.

\bibitem{tramer2016stealing}
Florian Tram{\`e}r, Fan Zhang, Ari Juels, Michael~K. Reiter, and Thomas
  Ristenpart.
\newblock Stealing machine learning models via prediction apis.
\newblock In {\em 25th {USENIX} Security Symposium ({USENIX} Security 16)},
  pages 601--618, Austin, TX, 2016. {USENIX} Association.

\bibitem{van2016drammer}
Victor Van Der~Veen, Yanick Fratantonio, Martina Lindorfer, Daniel Gruss,
  Cl{\'e}mentine Maurice, Giovanni Vigna, Herbert Bos, Kaveh Razavi, and
  Cristiano Giuffrida.
\newblock Drammer: Deterministic rowhammer attacks on mobile platforms.
\newblock In {\em Proceedings of the 2016 ACM SIGSAC conference on computer and
  communications security}, pages 1675--1689. ACM, 2016.

\bibitem{wang2018great}
Bolun Wang, Yuanshun Yao, Bimal Viswanath, Haitao Zheng, and Ben~Y. Zhao.
\newblock With great training comes great vulnerability: Practical attacks
  against transfer learning.
\newblock In {\em 27th {USENIX} Security Symposium ({USENIX} Security 18)},
  pages 1281--1297, Baltimore, MD, 2018. {USENIX} Association.

\bibitem{wang2018training}
Naigang Wang, Jungwook Choi, Daniel Brand, Chia-Yu Chen, and Kailash
  Gopalakrishnan.
\newblock Training deep neural networks with 8-bit floating point numbers.
\newblock In S.~Bengio, H.~Wallach, H.~Larochelle, K.~Grauman, N.~Cesa-Bianchi,
  and R.~Garnett, editors, {\em Advances in Neural Information Processing
  Systems 31}, pages 7675--7684. Curran Associates, Inc., 2018.

\bibitem{wu2016google}
Yonghui Wu, Mike Schuster, Zhifeng Chen, Quoc~V Le, Mohammad Norouzi, Wolfgang
  Macherey, Maxim Krikun, Yuan Cao, Qin Gao, Klaus Macherey, et~al.
\newblock Google's neural machine translation system: Bridging the gap between
  human and machine translation.
\newblock {\em arXiv preprint arXiv:1609.08144}, 2016.

\bibitem{xiao2013security}
Jidong Xiao, Zhang Xu, Hai Huang, and Haining Wang.
\newblock Security implications of memory deduplication in a virtualized
  environment.
\newblock In {\em 2013 43rd Annual IEEE/IFIP International Conference on
  Dependable Systems and Networks (DSN)}, pages 1--12. IEEE, 2013.

\bibitem{xiao2016one}
Yuan Xiao, Xiaokuan Zhang, Yinqian Zhang, and Radu Teodorescu.
\newblock One bit flips, one cloud flops: Cross-vm row hammer attacks and
  privilege escalation.
\newblock In {\em 25th {USENIX} Security Symposium ({USENIX} Security 16)},
  pages 19--35, Austin, TX, 2016. {USENIX} Association.

\bibitem{xu2015empirical}
Bing Xu, Naiyan Wang, Tianqi Chen, and Mu~Li.
\newblock Empirical evaluation of rectified activations in convolutional
  network, 2015.

\bibitem{zhou2018breaking}
Yan Zhou, Murat Kantarcioglu, and Bowei Xi.
\newblock Breaking transferability of adversarial samples with randomness,
  2018.

\end{thebibliography}
	
	\vspace{2.0em}
}

\begingroup				
\let\clearpage\relax	

\appendix
%
%


\renewcommand{\thesection}{\Alph{section}}


\section*{\centering{Appendix}}
\label{sec:appendix}

\section{Network Architectures}
\label{appendix:network-architectures}


\begin{table*}[ht]
\centering
\caption{\textbf{8 Network Architectures for MNIST.} We take the two baselines (Base and LeNet5) and make four and two variants from them, respectively. Note that we highlight the variations from the baselines in red color.}
\label{tbl:architecture-mnists}
%
\adjustbox{max width=0.9\textwidth}{
    \begin{tabular}{@{}cc|cc|cc|cc@{}}
    \toprule
    \multicolumn{2}{c|}{\textbf{Base}} & \multicolumn{2}{c|}{\textbf{Base (Wide)}} & \multicolumn{2}{c|}{\textbf{Base (Dropout)}} & \multicolumn{2}{c}{\textbf{Base (PReLU)}} \\ \midrule
    \textbf{Layer Type} & \textbf{Layer Size} & \textbf{Layer Type} & \textbf{Layer Size} & \textbf{Layer Type} & \textbf{Layer Size} & \textbf{Layer Type} & \textbf{Layer Size} \\ \midrule
    Conv (R) & 5x5x10 (2) & Conv (R) & 5x5x{\color{red}20} (2) & Conv (R) & 5x5x10 (2) & Conv ({\color{red}P}) & 5x5x10 (2) \\
    Conv (R) & 5x5x20 (2) & Conv (R) & 5x5x{\color{red}40} (2) & Conv (-) & 5x5x20 (2) & Conv ({\color{red}P}) & 5x5x20 (2) \\
    - & - & - & - & {\color{red}Dropout (R)} & {\color{red}0.5} & - & - \\
    FC (R) & 50 & FC (R) & \color{red}100 & FC (R) & 50 & FC ({\color{red}P}) & 50 \\
    - & - & - & - & {\color{red}Dropout (R)} & {\color{red}0.5} & - & - \\
    FC (S) & 10 & FC (S) & 10 & FC (S) & 10 & FC (S) & 10 \\ \midrule \midrule
    \multicolumn{2}{c|}{\textbf{Base (D-BNorm)}} & \multicolumn{2}{c|}{\textbf{LeNet5}~\cite{lecun2015lenet}} & \multicolumn{2}{c|}{\textbf{LeNet5 (Dropout)}} & \multicolumn{2}{c}{\textbf{LeNet5 (D-BNorm)}} \\ \midrule
    \textbf{Layer Type} & \textbf{Layer Size} & \textbf{Layer Type} & \textbf{Layer Size} & \textbf{Layer Type} & \textbf{Layer Size} & \textbf{Layer Type} & \textbf{Layer Size} \\ \midrule
    Conv (-) & 5x5x10 (2) & Conv (R) & 5x5x6 (2) & Conv (R) & 5x5x6 (2) & Conv (-) & 5x5x6 (2) \\
    {\color{red}BatchNorm (R)} & {\color{red}10} & - & - & - & - & {\color{red}BatchNorm (R)} & {\color{red}6} \\
     - & - & MaxPool (-) & 2x2 & MaxPool (-) & 2x2 & MaxPool (-) & 2x2 \\
    Conv (-) & 5x5x20 (2) & Conv (R) & 5x5x16 (2) & Conv (R) & 5x5x16 (2) & Conv (-) & 5x5x16 (2) \\
    {\color{red}BatchNorm (R)} & {\color{red}20} & - & - & - & - & {\color{red}BatchNorm (R)} & {\color{red}16} \\
    - & - & MaxPool (-) & 2x2 & MaxPool (-) & 2x2 & MaxPool (-) & 2x2 \\
    - & - & Conv (R) & 5x5x120 (2) & Conv (R) & 5x5x120 (2) & Conv (R) & 5x5x120 (2) \\
    - & - & - & - & - & - & {\color{red}BatchNorm (R)} & {\color{red}120} \\
    
    {\color{red}Dropout (R)} & {\color{red}0.5} & - & - & {\color{red}Dropout (R)} & {\color{red}0.5} & {\color{red}Dropout (R)} & {\color{red}0.5} \\
    - & - & MaxPool (-) & 2x2 & MaxPool (-) & 2x2 & MaxPool (-) & 2x2 \\
    - & - & Conv (R) & 5x5x120 (1) & Conv (R) & 5x5x120 (1) & Conv (R) & 5x5x120 (1) \\
    FC (R) & 50 & FC (R) & 84 & FC (R) & 84 & FC (R) & 84 \\
    {\color{red}Dropout (R)} & {\color{red}0.5} & - & - & {\color{red}Dropout (R)} & {\color{red}0.5} & {\color{red}Dropout (R)} & {\color{red}0.5} \\
    FC (S) & 10 & FC (S) & 10 & FC (S) & 10 & FC (S) & 10 \\ \bottomrule
    \end{tabular}
}
\end{table*}

We use 19 DNN models in our experiments: six architecture and their variants.
Table~\ref{tbl:architecture-mnists} 
describes two architectures and their six variations for MNIST.
For CIFAR10, we employ the base architecture from~\cite{suciu2018fail} that has four convolutional layers and a fully-connected layer, and we make three variations of it.
CIFAR10-AlexNet\footnote{\scriptsize\url{https://github.com/bearpaw/pytorch-classification/blob/master/models/cifar/alexnet.py}} and CIFAR10-VGG16\footnote{\scriptsize\url{https://github.com/kuangliu/pytorch-cifar/blob/master/models/vgg.py}} are from the community.
For ImageNet, we use the DNN architectures available from the Internet\footnote{\scriptsize\url{https://github.com/pytorch/vision/tree/master/torchvision/models}}.
In Sec~\ref{subsec:low-precisions}, we employ two networks (8-bit quantized\footnote{\scriptsize\url{https://github.com/eladhoffer/quantized.pytorch}} and binarized versions of MNIST-L5) from the community\footnote{\scriptsize\url{https://github.com/jiecaoyu/XNOR-Net-PyTorch}} with adjustments.

\section{The Vulnerability Using Different Criteria}
\label{append:rads-with-single-bitflips}

We examine the vulnerable parameter ratio (vulnerability) using the different RAD criterion with 15 DNN models.
Our results are in Figure~\ref{fig:damage-plots}.
Each figure describe the vulnerable parameter ratio on a specific RAD criterion; for instance, in MNIST-L5, the model has 40\% of vulnerable parameters that cause [RAD$>0.5$], which estimates the upper bound of the \MakeLowercase{\prob} attacker.
In MNIST, CIFAR10, and two ImageNet models, the vulnerability decreases as the attacker aims to inflict the severe damage; however, in ImageNet, ResNet50, DenseNet161, and InceptionV3 have almost the same vulnerability ($\sim$50\%) with the high criterion [RAD$>0.8$].

\begin{figure*}[h]
	\centering
	\includegraphics[width=0.32\linewidth]{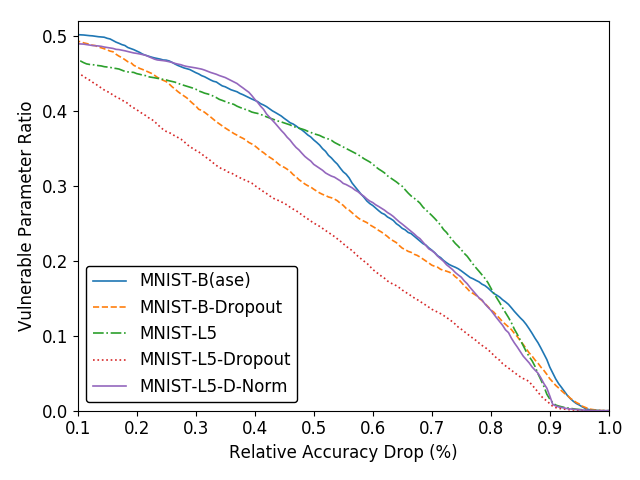}
	\includegraphics[width=0.32\linewidth]{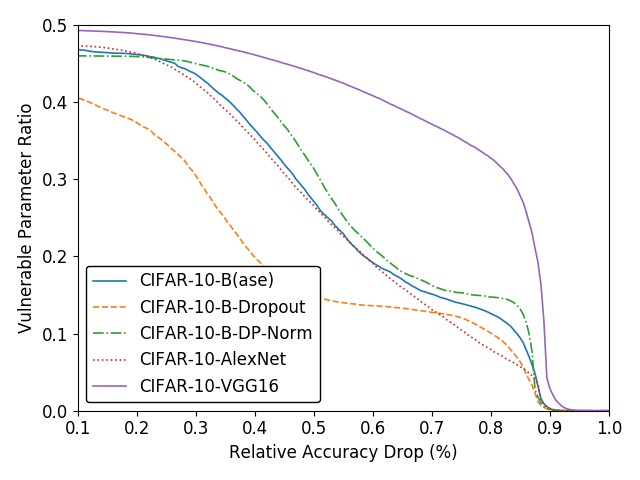}
	\includegraphics[width=0.32\linewidth]{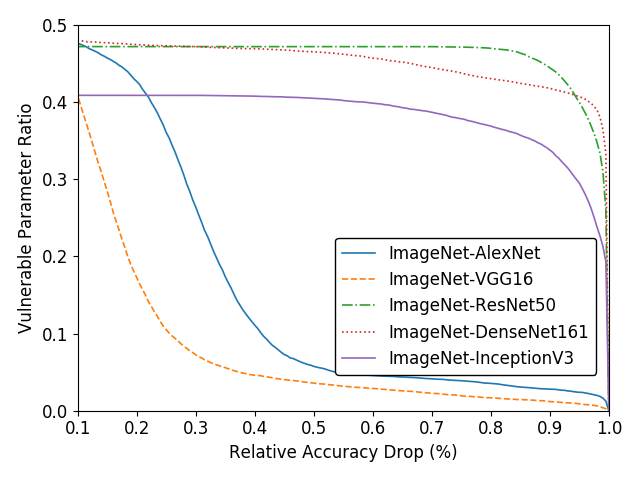}
	\caption{\textbf{The vulnerability of 15 DNN models using different criteria.} We plot the vulnerable parameter ratio based on the different RADs that an attacker aims; 5 from MNIST (left), 5 from CIFAR10 (middle), and 5 from ImageNets (right).}
	\label{fig:damage-plots}
	\vspace{-1.0em}				
\end{figure*}

\section{Hyper-parameters for Training}
\label{appendix:dataset-stats-hyperparams}

In our experiments, we use these hyper-parameters:
\begin{itemize}
	\item \textbf{MNISTs.} For 
	MNIST models, we use
	: SGD, 40 epochs, 0.01 learning rate (lr), 64 batch, 0.1 momentum, and adjust learning rate by 0.1, in every 10 epochs.
	\item \textbf{CIFAR10s.} For Base models 
	we use
	: SGD, 50 epochs, 0.02 lr, 32 batch, 0.1 momentum, and adjust lr by 0.5, in every 10 epochs. For AlexNet, we use: 
	300 epochs, 0.01 lr, 64 batch, 0.1 momentum, and adjust lr by 0.95, in every 10 epochs. For VGG16, we use: 
	300 epochs, 0.01 lr, 128 batch, 0.1 momentum, and adjust lr by 0.15, in every 100 epochs.
	\item \textbf{GTSRB.} 
	We fine-tune 
	VGG16 pre-trained on ImageNet, using: SGD, 40 epochs, 0.01 lr, 32 batch, 0.1 momentum, and adjust lr by 0.1 and 0.05, in 15 and 25 epochs. We freeze the parameters of the first 10 layers.
	\item \textbf{Flower102.} 
	We fine-tune 
	ResNet50 pre-trained on ImageNet, using: SGD, 40 epochs, 0.01 lr, 50 batch, 0.1 momentum, and adjust lr by 0.1, in 15 and 25 epochs. We freeze the parameters of the first 10 layers.
\end{itemize}
\textcolor{white}{.}

\endgroup

\end{document}